\documentclass[12pt]{article}

\usepackage{graphicx,rotating,epsfig,amsmath,amssymb,color,multirow,pstricks,pst-plot,comment}
\newlength{\digitwidth} 

\newcommand{\jpsi}{\mbox{J/$\psi$}}

\newcommand{\psip}{\mbox{$\psi(2S)$}}

\newcommand{\chicOne}{\mbox{$\chi_{c1}$}}
\newcommand{\chicTwo}{\mbox{$\chi_{c2}$}}

\newcommand{\upsAllS}{\mbox{$\Upsilon(nS)$}}
\newcommand{\upsOneS}{\mbox{$\Upsilon(1S)$}}
\newcommand{\upsTwoS}{\mbox{$\Upsilon(2S)$}}
\newcommand{\upsThreeS}{\mbox{$\Upsilon(3S)$}}

\newcommand{\pt}{\mbox{$p_{\rm T}$}}

\newcommand{\QQbar}{\mbox{$Q \overline{Q}$}}

\newcommand{\cinst}[2]{$^{\mathrm{#1)}}$~#2\par}
\newcommand{\crefi}[1]{$^{\mathrm{#1)}}$}
\newcommand{\HRule}{\rule{0.4\linewidth}{0.3mm}}

\begin{document}

\begingroup
\thispagestyle{empty} \baselineskip=14pt
\parskip 0pt plus 5pt

\begin{center}
{\large EUROPEAN LABORATORY FOR PARTICLE PHYSICS}
\end{center}

\bigskip
\begin{flushright}
~\\ CERN--PH\\
~  March 11, 2014
\end{flushright}

\bigskip
\begin{center}
{\Large\bf
Quarkonium production in the LHC era:\\[3mm]
a polarized perspective}

\bigskip\bigskip

Pietro Faccioli\crefi{1,2},
Valentin Kn\"unz\crefi{3},
Carlos Louren\c{c}o\crefi{4},\\
Jo\~ao Seixas\crefi{1,2}
and Hermine K.\ W\"ohri\crefi{4}

\bigskip\bigskip\bigskip
\textbf{Abstract}

\end{center}

\begingroup
\leftskip=0.4cm \rightskip=0.4cm \parindent=0.pt

Polarization measurements are usually considered as the most difficult challenge for the 
QCD description of quarkonium production.
In fact, global data fits for the determination of the non-perturbative parameters of 
bound-state formation traditionally exclude polarization observables and use them as 
a posteriori verifications of the predictions, with perplexing results.
With a change of perspective, we move polarization data to the centre of the study, 
advocating that they actually provide the strongest fundamental indications about the 
production mechanisms, even before we explicitly consider perturbative calculations.

Considering \psip\ and \upsThreeS\ measurements from LHC experiments and 
state-of-the-art NLO short-distance calculations in the framework of non-relativistic QCD 
factorization (NRQCD), we perform a search for a kinematic domain where the polarizations 
can be correctly reproduced together with the cross sections, by systematically scanning 
the phase space and accurately treating the experimental uncertainties.
This strategy provides a straightforward solution to the ``quarkonium polarization puzzle'' 
and reassuring signs that the theoretical framework is reliable. At the same time, the 
results expose unexpected hierarchies in the non-perturbative NRQCD parameters, 
that open new paths towards the understanding of bound-state formation in QCD.

\bigskip
\endgroup
\vspace{1cm}
\begin{center}
\emph{Submitted to Phys.\ Lett.\ B}
\end{center}
\vfill
\begin{flushleft}
\HRule\\
\small
\cinst{1} {Laborat\'orio de Instrumenta\c{c}\~ao e F\'{\i}sica Experimental de
  Part\'{\i}culas (LIP), Lisbon, Portugal}
\cinst{2} {Physics Department, Instituto Superior T\'ecnico (IST), Lisbon, Portugal}
\cinst{3} {Institute of High Energy Physics (HEPHY), Vienna, Austria}
\cinst{4} {European Organization for Nuclear Research (CERN), Geneva, Switzerland}
\end{flushleft}
\endgroup

\newpage
\sloppy

\section{Introduction, motivation and conceptual remarks}
\label{sec:intro}

Up to the early 1990's, quarkonium production was believed to be reasonably well
described by the (leading order, LO) colour singlet model (CSM)~\cite{bib:Baier},
which basically assumes that the produced quark-antiquark (\QQbar)
pair has, since its inception, the spin ($S$), angular momentum
($L$) and colour quantum numbers of the observable quarkonium, and that no
$L$/$S$-changing transitions occur during the bound-state formation.
Measurements of the \jpsi\ and \psip\ production cross sections by the
E789 fixed-target experiment at Fermilab~\cite{bib:E789}, 
which exceeded the predicted values by factors of 7 and 25, respectively,
challenged this model. Nonetheless, since the data were collected
at relatively low transverse momentum, \pt, where non-perturbative effects not addressed
by the CSM could dominate, this discrepancy was not immediately
perceived as a potentially serious problem.

The results obtained by CDF at the Tevatron~\cite{bib:CDFpsisRunI}
covered a much higher \pt\ range and showed an even larger difference between
the calculated and measured prompt production cross sections 
(after subtracting the non-negligible yield from B meson decays).
The \jpsi\ discrepancy could be attributed to feed-down contributions from 
decays of the \mbox{$P$-wave} states $\chi_{c1}$ and $\chi_{c2}$,
poorly-known experimentally and expected to dominate prompt \jpsi\
production because the directly-produced \jpsi\ component was supposed to
be phase-space suppressed by the need of an extra gluon.
The clear-cut perception that there was a problem in the understanding of
directly-produced charmonium production was unravelled by the \psip\ 
measurement, insensitive to feed-down decays and a factor 50 higher than
calculated in the CSM, a surprisingly large discrepancy nicknamed 
``the CDF \psip\ anomaly''.

Meanwhile, significant progress was being made on the theory side,
with the birth of the non-relativistic quantum chromodynamics
(NRQCD) factorization approach~\cite{bib:NRQCD}. While in the CSM
the observed bound-state meson can only result from quark pairs produced 
in a singlet state, NRQCD includes terms where the original quark pairs are 
in colour-octet states.
In this effective field
theory, the non-perturbative evolution that converts the (coloured)
\QQbar\ into a physically-observable bound meson, possibly changing
$L$ and/or $S$, is described by long-distance matrix elements
(LDMEs), factorized from the parton-level contributions. In NRQCD
the LDMEs are supposed to be constant (i.e.\ independent of the
\QQbar\ momentum) and universal (i.e.\ process-independent). They
cannot be calculated within the theoretical framework and need to
be determined by comparisons to experimental data.

The NRQCD formalism was greeted with encouraging words given its
seeming success in reproducing the CDF charmonia \pt-differential cross 
sections.
One should bear in mind, however, that the colour-octet terms
have free normalisations (essentially determined by the LDMEs), so that 
we should not be very surprised if the measured cross sections can be 
well described (the CSM, instead, had zero adjustable parameters).

A very reasonable way of evaluating if a given theory provides a suitable 
representation of reality is to fix its free parameters through fits to 
a given set of measurements and then check how well it predicts other 
physical observables, not previously considered. This procedure has been
followed over the last years in quarkonium production studies:
first the NRQCD LDMEs are determined by fitting the cross-section
measurements to a superposition of singlet and octet terms; then the
resulting model is used to predict the quarkonium polarizations. The
outcome (see Ref.~\cite{bib:QWGreportII} and references therein)
is that the predicted quarkonium
polarizations are very different from the measured ones, a situation
dubbed ``the quarkonium polarization puzzle".

It is worth emphasizing at this point that the detailed NRQCD modelling 
of quarkonium production crucially depends on the LDMEs, which are
determined by the quality (and variety) of the
available experimental measurements. This, by itself, is not an
uncommon situation in high-energy physics. For instance, most QCD
calculations require the use of data-driven parton distribution
functions (PDFs) and fragmentation functions (FFs) to translate
parton-level calculations into predictions that can be compared to
experimental (hadron-level) data. Assuming that the PDFs and FFs are
universal, one can measure them using suitable processes (like
deep-inelastic scattering and $e^+e^-$ interactions) and then use
them as inputs for calculating other measured processes.

But why is it that, in past quarkonium production studies, only the
cross-section measurements were used to fit the NRQCD LDMEs? 
This is obviously not the only viable strategy. One could start by
fitting the polarization measurements and then predict the
differential cross sections, apart from their absolute normalizations;
or, more democratically, one could make a global fit of both sets of
measurements, cross sections and polarizations.
The answer is that quarkonium polarization measurements are very
complex and require exceptional care in the corresponding data
analyses. Most of the quarkonium polarization results published
before 2011 are incomplete and ambiguous~\cite{bib:EPJC69}.
Results obtained by the CDF and D0 Tevatron experiments, in particular, 
have been plagued by a series of suspicious observations, with at least
two cases (CDF Run~1 versus CDF Run~2 for the 
\jpsi~\cite{bib:CDFpol1, bib:CDFpol2} and CDF versus D0 for 
the \upsOneS~\cite{bib:CDFUpsPolRunI,bib:D0UpsPolRunII}) 
where two measurements mutually excluded each other. 
In these conditions, it is not surprising to see the fundamental role of 
polarization measurements purposely downgraded to an 
\emph{a posteriori} crosscheck of the predictions.

The experimental situation has dramatically improved in the last 
years, first with the most recent CDF measurement of the three
\upsAllS\ polarizations~\cite{bib:CDFUpsPolRunII}, 
and then with the LHC measurements, made by CMS for the five 
\mbox{$S$-wave} quarkonium states 
(two charmonia~\cite{bib:CMSpsiPol} and 
three bottomonia~\cite{bib:CMSUpsPol}) 
and by LHCb for the \jpsi~\cite{bib:LHCbpsiPol}. 
All of these studies were made following the much-improved
methodologies proposed in a series of recent 
publications~\cite{bib:EPJC69, bib:FT2C, bib:LTGen, bib:ImprovedQQbarPol}. 
The good mutual consistency of all these recent
results reflects their vastly improved robustness with respect to
the previous measurements.

These new results allow for a change of strategy. We can now proceed
with ``global fits'' of quarkonium data, considering the
polarization measurements at the same level as the cross sections.
Actually, polarization is much more straightforwardly related to the
variables of the theory than the momentum distributions, the
different colour channels for \QQbar\ production being characterized
by simple and distinctive polarization patterns. This consideration
guides our analysis, allowing us to make immediate simplifications
and improve the robustness of the results.

To fully benefit from the improved quality and constraining power of
the new measurements, efforts must be devoted to a careful treatment
of the experimental and theoretical uncertainties. Correlations
induced by systematic uncertainties due to luminosity measurements
must be correctly taken into account. A more complex type of
correlation is the one induced by the strong dependence of the
acceptance determinations on the shape of the dilepton decay
distributions. Previously-reported NRQCD global fits 
use differential cross sections measured with acceptance corrections
evaluated assuming unpolarized production, ignoring the large
uncertainty that the experiments assign to the lack of prior
knowledge about the quarkonium polarization. This imprecision
must be corrected, at least to ensure logical consistency in cases
leading to predictions of significantly-polarized
quarkonia. As a further improvement, theoretical uncertainties must
be modeled in the fitting algorithm, in order to allow for a
realistic evaluation of the goodness of fit. This
crucial aspect of the verification of the theory has not been
properly addressed in previous analyses.

Furthermore, we should also revisit the very spirit and motivation
of these fits. In past studies, measurements made at rather low \pt,
even lower than the mass of the quarkonium state, have been included
in the NRQCD fits. At first sight, this might seem a good idea,
because the lowest \pt\ data points are usually the ones with the
best statistical accuracy and, hence, are the ones that most strongly
constrain the free parameters of the fits. However, the NRQCD
\emph{factorization} approach~\cite{bib:NRQCD} 
requires that the short-distance and
long-distance phases of the quarkonium production process can be
factorized, i.e.\ there must be a sharp separation between the typical 
\QQbar\ distance scales of the hard scattering process, 
$\sim \max(1/ p_{\rm T}, 1/m_{Q})$, 
and of the bound-state, $\sim 1/(m_{Q}v)$, 
where $m_{Q}$ is the heavy-quark mass 
and $v$ its velocity in the \QQbar\ rest frame.
Effectively, this means that we cannot expect that the NRQCD calculations
reproduce the measurements at low \pt\
(especially given that the presently available calculations are limited to 
next-to-leading order in $\alpha_s$, NLO)
and, hence, using those data to constrain the fitted parameters is a priori unjustified.
Even worse, their high
statistical accuracy might lead the fit into strongly biased results.
When we compare data to theory, the pertinent question is not 
``Is NRQCD a valid theory for heavy quarkonium production?''
but rather 
``Is there a kinematic domain in which NRQCD is a
valid theory for heavy quarkonium production?''. To search for
possible domains of validity of NRQCD factorization, 
at the present status of the perturbative calculations, in the
description of charmonium and bottomonium production, we will apply
progressively changing kinematic thresholds to the data and study
the corresponding variations in the fit results.

The paper is organized as follows. After a preliminary description
of the theory ingredients in Section~\ref{sec:definitions}, we present
in Section~\ref{sec:indications} some basic considerations that drive
our analysis, inspired by recurring patterns in the data and in
particular in the polarization measurements. The study of the
possible domain of validity of NLO NRQCD factorization based on LHC data
is described in Section~\ref{sec:results_vs_pTmin}. The main results of
the global analysis are presented in Section~\ref{sec:results_best} and
discussed in Section~\ref{sec:discussion}.

\section{Theory ingredients}
\label{sec:definitions}

\begin{figure}[h!]
\centering
\includegraphics[width=0.55\linewidth]{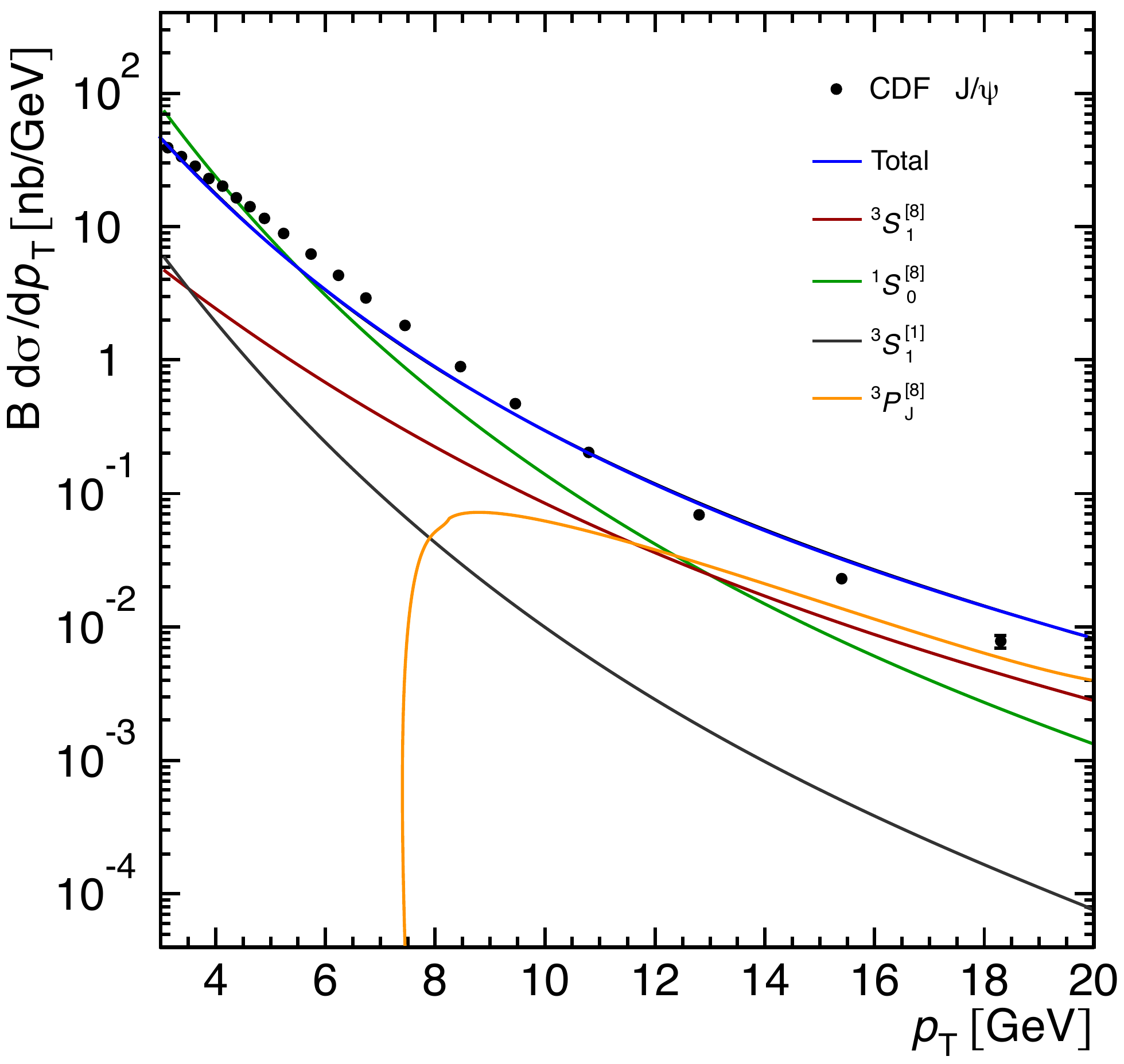}
\caption{Individual SDCs calculated at NLO~\cite{bib:BKmodel} 
for \jpsi\ production,
scaled by LDMEs fitted~\cite{bib:BKNPB,bib:BKMPLA} 
to the CDF data~\cite{bib:CDFjpsi2005}.}
\label{fig:SDCatNLO}
\end{figure}

In the hypothesis of factorization of short- and long-distance
effects, the cross section for the inclusive production of the bound
meson $H$ (plus unobserved particles, $X$) in a collision of initial
systems $A$ and $B$ is expressed by the formula
\begin{equation}
  \sigma(A+B \to H+X) = \sum\limits_{S,L,C}
  \mathcal{S}(A+B\to \QQbar[^{2S+1}L_J^C] +X) \times
  \mathcal{L}( \QQbar[^{2S+1}L_J^C] \to H)  \; .
\label{eq:factorization_formula}
\end{equation}
%
%
In each term of the sum, the (kinematics-dependent) short-distance
coefficient (SDC), $\mathcal{S}$, is proportional to the
parton-level cross section for the production of the pre-resonance
$\QQbar$ in a given angular momentum ($L$,$S$) and colour ($C$)
configuration, while the corresponding (constant)
LDME, $\mathcal{L}$, is proportional to the probability of the bound-state
formation. The theory ingredients of our analysis of \psip\ and
\upsThreeS\ production in pp collisions at the LHC are the
perturbative calculations of the SDCs for colour-octet and
colour-singlet \QQbar\ pairs, and the corresponding polarizations.
The LDMEs are fit parameters determined from data. According to
NRQCD $v$-scaling rules, the dominating contributions to the
production of \mbox{$S$-wave} vector quarkonia are the colour-singlet
($^3S_1^{[1]}$) and three colour-octet ($^1S_0^{[8]}$, $^3S_1^{[8]}$
and $^3P_J^{[8]}$) channels. We use the calculations made at NLO
reported in Ref.~\cite{bib:BKmodel}, provided for a rest
energy of the colour-singlet or colour-octet pre-resonance
state $E_{Q \overline{Q}} = 3$~GeV. Figure~\ref{fig:SDCatNLO}
shows the individual contributions (products of \pt-dependent SDCs
times constant LDMEs) of the four \QQbar\ colour configurations, and
their sum, compared to the 
\jpsi\ cross section measured by CDF~\cite{bib:CDFjpsi2005}.
The LDMEs multiplying the octet SDCs have
been obtained from a global fit of hadro- and photo-production
data~\cite{bib:BKNPB,bib:BKMPLA}.

\begin{figure}[h!]
\centering
\includegraphics[width=0.55\linewidth]{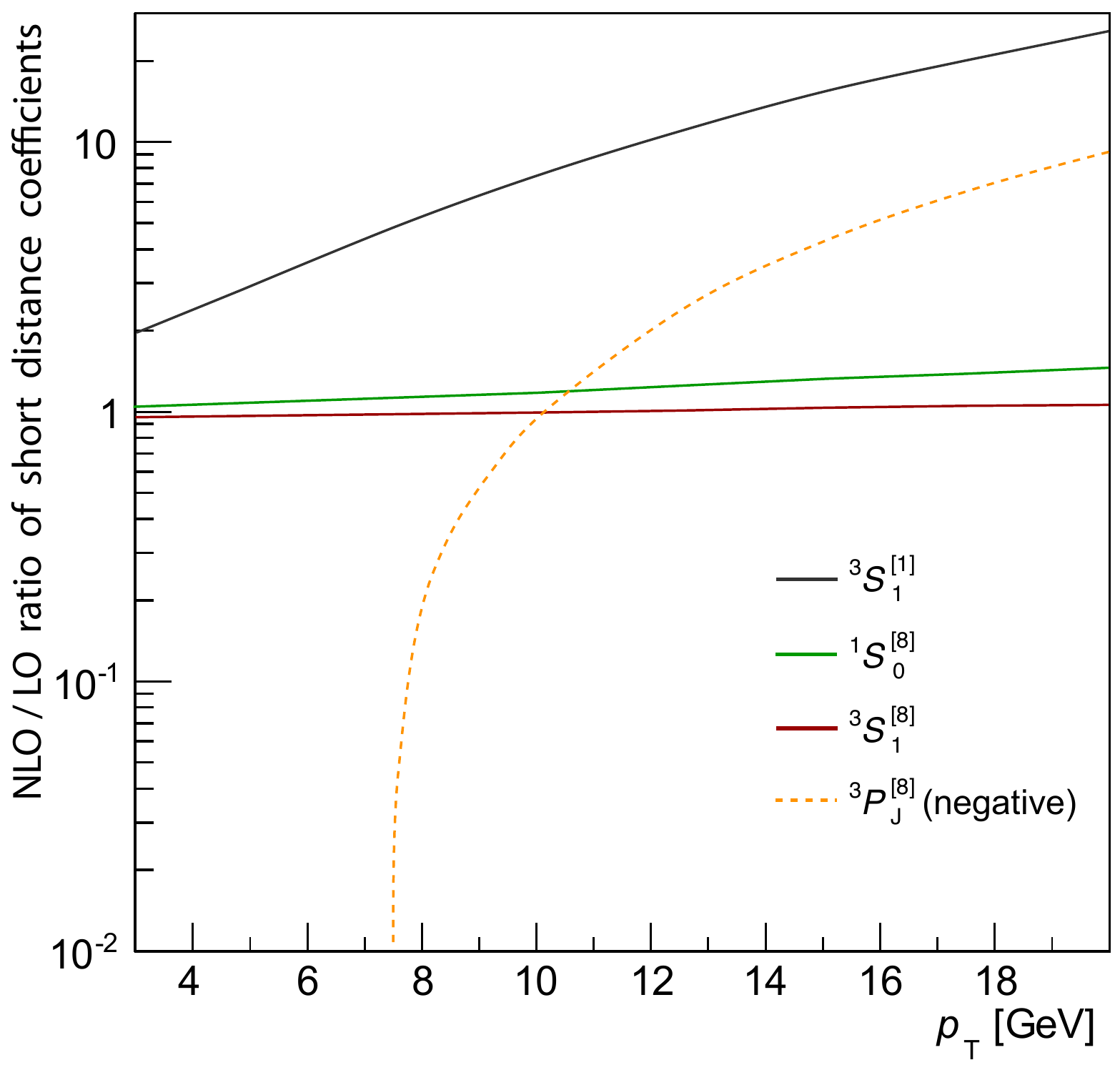}
\caption{Ratios between the NLO and LO SDCs~\cite{bib:BKmodel}.}
\label{fig:SDCratios}
\end{figure}

\begin{figure}[ht!]
\centering
\includegraphics[width=0.65\linewidth]{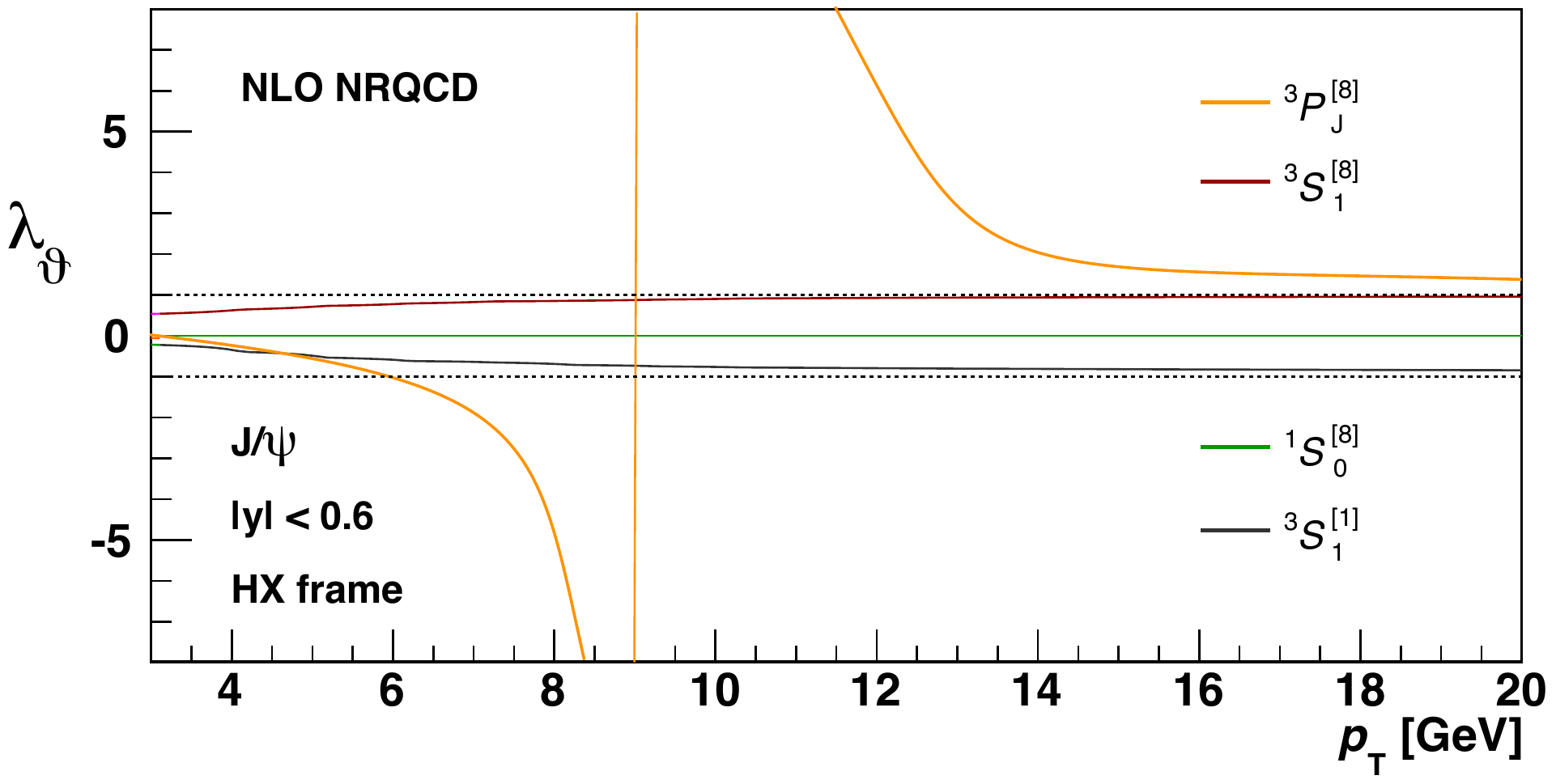}
\caption{Polarizations parameters, $\lambda_\vartheta$, calculated at NLO
for the colour-singlet term and for the three colour-octet terms~\cite{bib:BKmodel}.}
\label{fig:individualPols}
\end{figure}

Figure~\ref{fig:SDCratios}
illustrates how the individual SDCs change from LO to NLO.
It is worth noting that the $^3P_J^{[8]}$ SDC, for \pt\ above 7.5~GeV,
changes from positive at LO to negative at NLO.
Figure~\ref{fig:individualPols} shows the \pt\ dependence of the 
polarization parameters $\lambda_\vartheta$, calculated at NLO
for vector quarkonia produced in different \QQbar\ 
colour configurations, where $\lambda_\vartheta = (\mathcal{S}_{\rm
T}-\mathcal{S}_{\rm L})/(\mathcal{S}_{\rm T}+\mathcal{S}_{\rm L})$ 
and $\mathcal{S}_{\rm T}$ ($\mathcal{S}_{\rm L}$) is the
transverse (longitudinal) short distance cross section,
in the helicity frame (HX).
At LO, except for small deviations at low \pt, vector quarkonia
have $\lambda_\vartheta$ either equal to $+1$ 
(from $^3S_1^{[1]}$ or $^3S_1^{[8]}$) 
or to $0$ (from $^1S_0^{[8]}$ or $^3P_J^{[8]}$).

Among the colour-octet contributions, the $^3P_J^{[8]}$
short-distance cross section raises attention for several
peculiarities, calling for extra efforts in improved calculations. 
Firstly, both the \QQbar\ yield and its polarization
change drastically from LO to NLO. Secondly, at NLO they
have unphysical behaviours, the yield 
being negative at low or high \pt, depending on the sign of 
the corresponding LDME,
and the polarization parameter $\lambda_\vartheta$ reaching
values higher than $+1$ (and even diverging for a certain value of
\pt, when $\mathcal{S}_{\rm T} = -\mathcal{S}_{\rm L}$).
One might be tempted to argue that this is not 
a conceptual problem, since the cross section and polarization 
are not observable for each individual subprocess; only the sum 
over all subprocesses must lead to physically-meaningful observables.
However, at least in principle and at least in some phase space corner, 
the exact cancellation of the unphysical effects may be affected by
approximations in the models (including the use of the model outside
its domain of validity), by a not sufficiently accurate treatment of
the experimental constraints on the theoretical parameters, or even
simply by their statistical/systematic fluctuations. Fits relying on
delicate compensations clearly demand special care.

\section{Data-driven considerations}
\label{sec:indications}

\begin{figure}[h]
\centering
\includegraphics[width=0.67\linewidth]{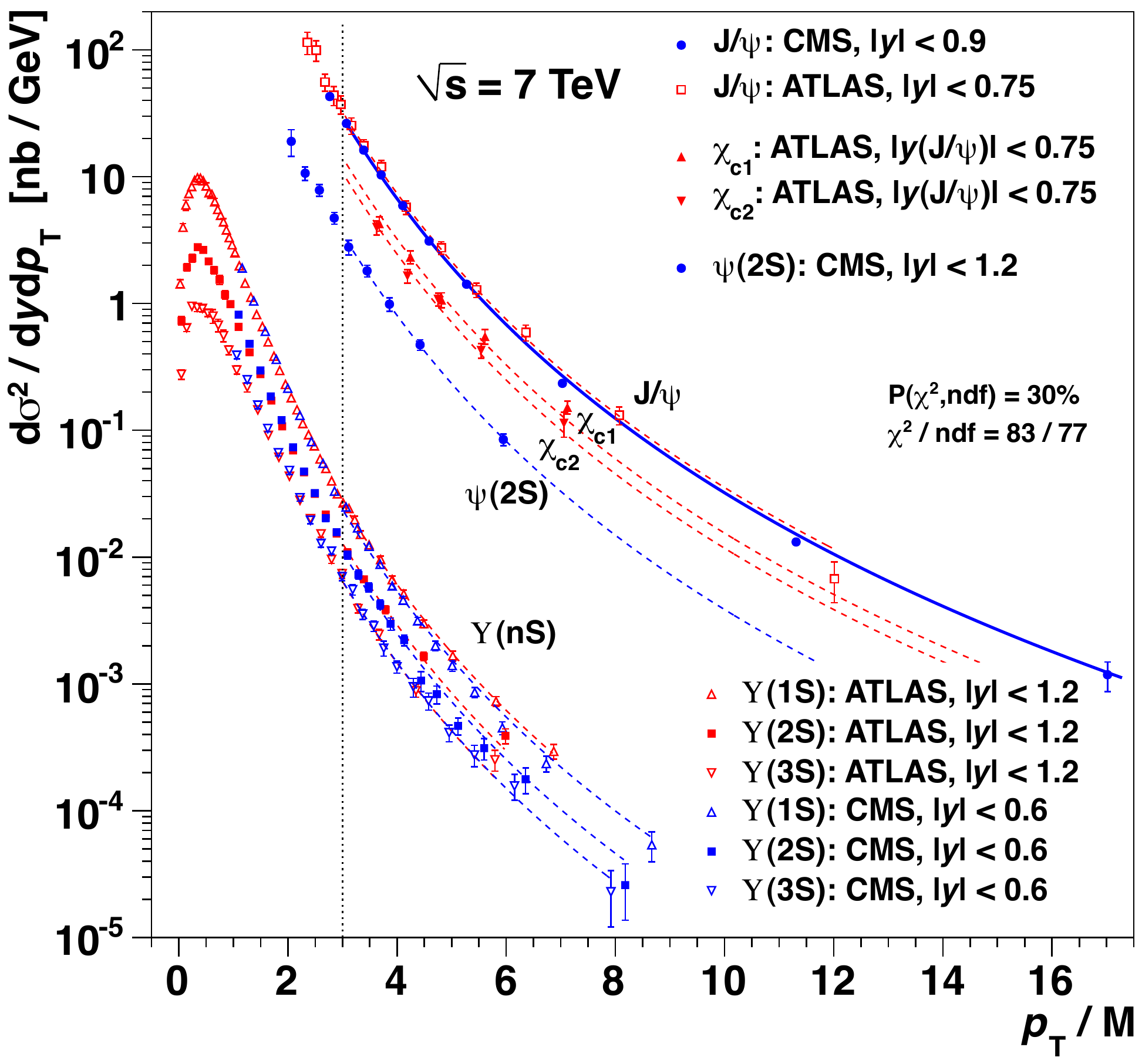}
\caption{Mid-rapidity quarkonium $p_{\rm{T}}/M$ distributions measured at 
$\sqrt{s} = 7$~TeV by the ATLAS and CMS 
experiments~\cite{bib:ATLASjpsi2011, bib:CMSjpsi2012,
bib:ATLASYnS2013, bib:CMSYnS2013, bib:ATLASchi2013}.
The solid curve is a fit to the \jpsi\ data of CMS (for $p_{\rm{T}}/M>3$), 
using a power-law function~\cite{bib:HERAb}, 
while the dashed curves are replicas 
with normalizations adjusted to the individual datasets.}
\label{fig:pToverMscaling}
\end{figure}

Our analysis is inspired and guided by two main data-driven
considerations. The first is illustrated by
Fig.~\ref{fig:pToverMscaling}, which shows the differential cross
sections for the production of seven different quarkonium states, as
measured by the ATLAS and CMS 
experiments~\cite{bib:ATLASjpsi2011, bib:CMSjpsi2012,
bib:ATLASYnS2013, bib:CMSYnS2013, bib:ATLASchi2013}. 
We applied a mass rescaling
to the \pt\ variable in order to equalize the kinematic effects of
different average parton momenta and phase spaces. 
When transformed to $p_{\rm{T}}/M$ distributions, the shapes of the 
differential cross sections of these seven states are well described 
(at least for $p_{\rm{T}}/M > 3$) by a simple empirical 
function~\cite{bib:HERAb}, with common values of its two shape 
parameters (the normalized $\chi^2$ of the global fit is 1.1 with 77 
degrees of freedom).

The easiest conjecture explaining this common behaviour is that a
very simple composition of processes, probably dominated by one
single mechanism, is responsible for the production of
all quarkonia. If several mechanisms were simultaneously at
play, we would expect to see variations of their mixture because
the differences in the masses of the component quarks and in the
binding energy of the observed hadrons should induce changes in the
non-perturbative effects.
We must also keep in mind that the production kinematics
addressed by these measurements differ from each other in that they
contain almost pure \mbox{$S$-wave} (\psip\ and \upsThreeS) 
or \mbox{$P$-wave} ($\chi_{c1}$ and $\chi_{c2}$)
contributions or, because of feed-down effects, a mixture of the two
(\jpsi, \upsOneS, \upsTwoS).
If confirmed with higher
precision, the observed $p_{\rm{T}}/M$ scaling would provide a strong
physical indication without relying on explicit theoretical calculations. 
In fact, since the kinematics of colour-singlet processes is necessarily 
dependent on the angular momentum of the observed state, observing
that states of different angular momentum quantum numbers are
produced perturbatively with identical kinematics directly implies that 
colour-singlet processes play a negligible role.

\begin{figure}[h]
\centering
\includegraphics[width=0.48\linewidth]{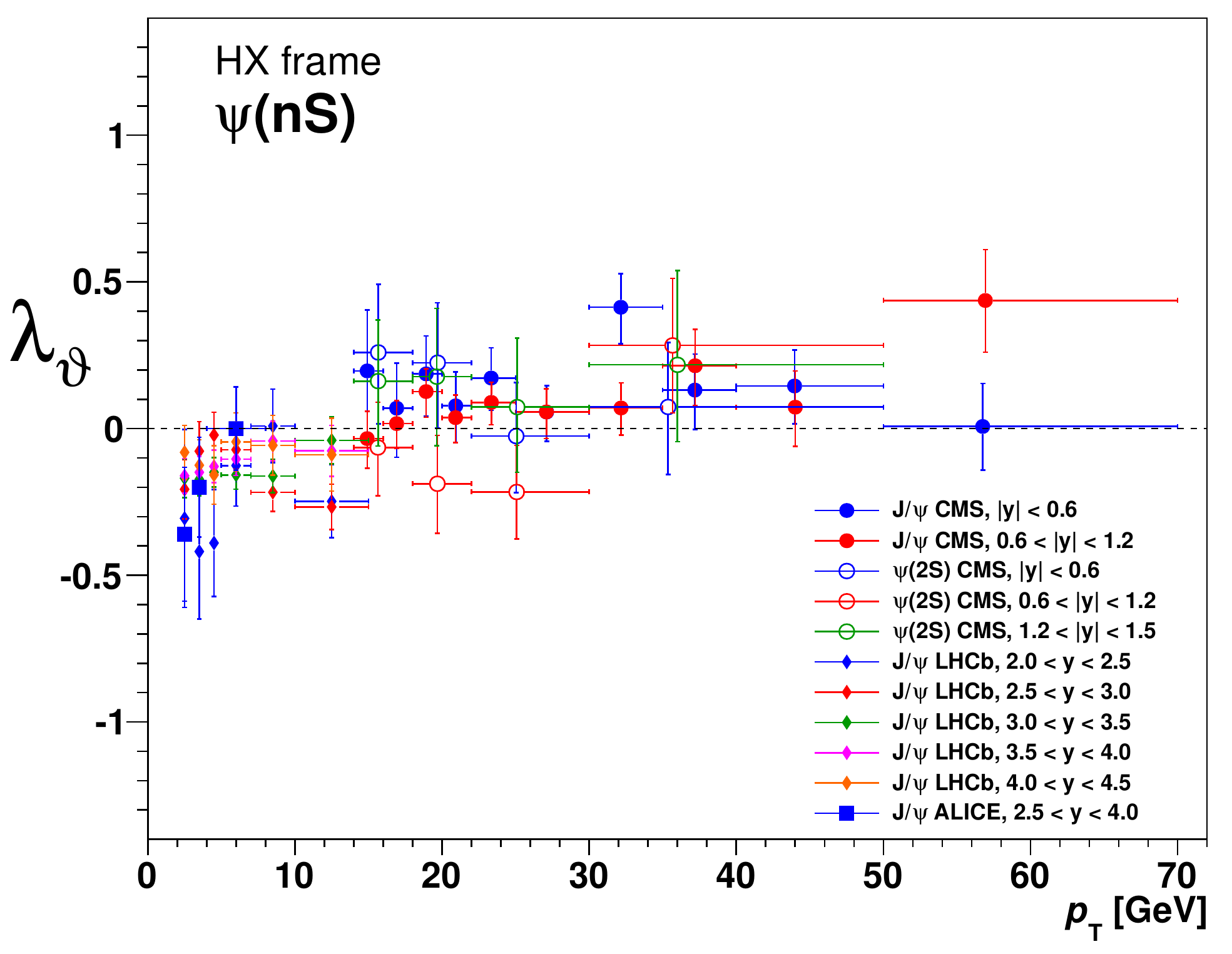}
\includegraphics[width=0.48\linewidth]{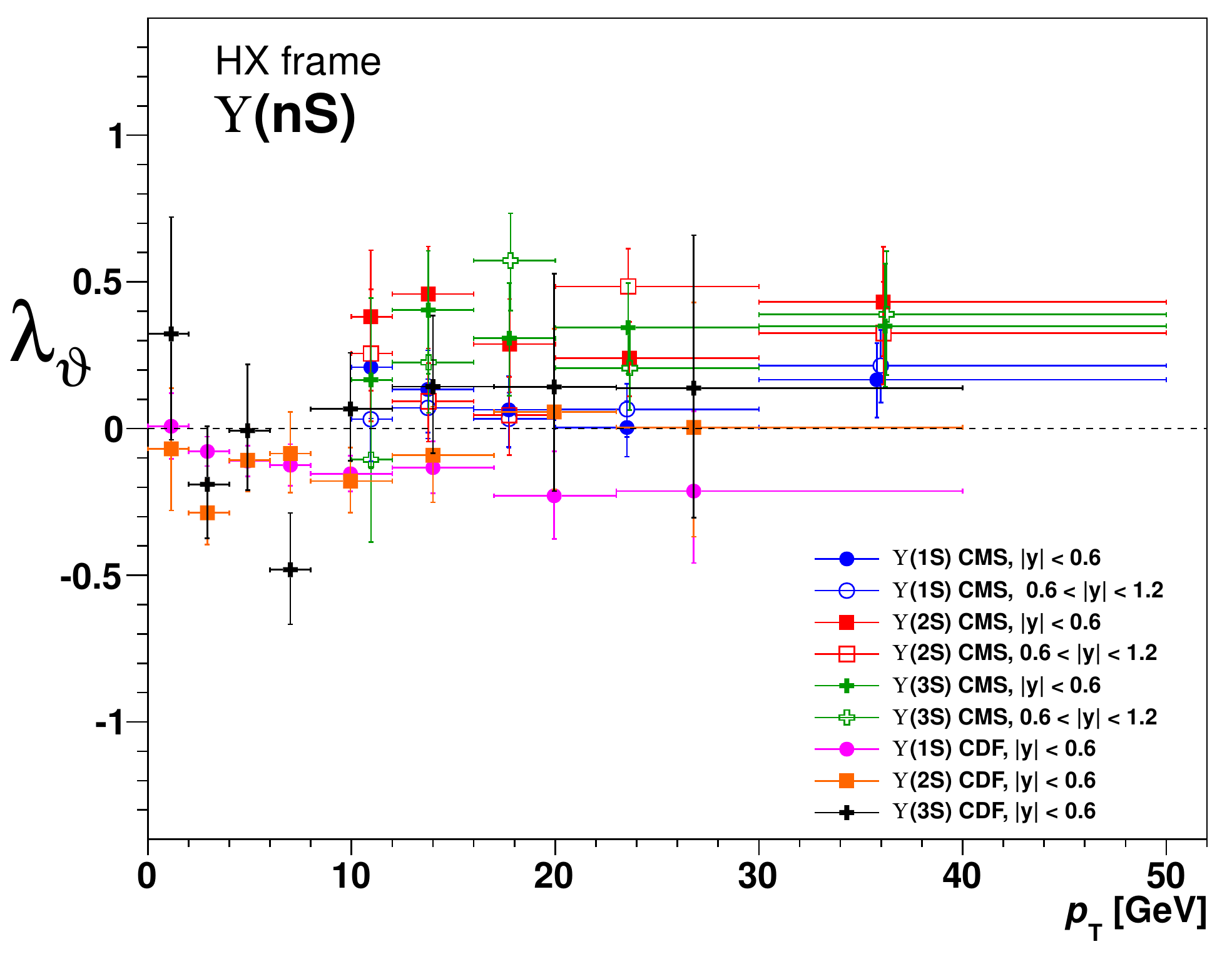}
\caption{Charmonium (top) and bottomonium (bottom) polarizations, 
as measured by 
CDF~\cite{bib:CDFUpsPolRunII}, 
ALICE~\cite{bib:ALICEpol}, 
CMS~\cite{bib:CMSpsiPol,bib:CMSUpsPol} and
LHCb~\cite{bib:LHCbpsiPol}.}
\label{fig:polMeasurements}
\end{figure}

The second, even stronger, hint comes from the quarkonium polarization
measurements. As shown in Fig.~\ref{fig:polMeasurements}, the
polarizations of the \mbox{$S$-wave} quarkonia recently measured by
CDF~\cite{bib:CDFUpsPolRunII} and at the 
LHC~\cite{bib:ALICEpol, bib:CMSpsiPol, bib:CMSUpsPol, 
bib:LHCbpsiPol} cluster around the unpolarized
limit, with no significant dependencies on \pt\ or rapidity, no
strong changes from directly-produced states to those affected by
\mbox{$P$-wave} feed-down decays, and no evident differences between
charmonium and bottomonium. This observation strengthens the
conjecture that, in ``zero-order'' approximation, all quarkonia are
dominantly produced by a single mechanism. 
Naturally, the polarization observable
has an immediate interpretation in terms of angular momentum
properties, especially strong given the peculiarity of the
unpolarized result: the dominating channel must be the
one leading to the ``ground-state'' pre-resonance object
$^1S_0^{[8]}$.

\begin{figure}[tb]
\centering
\includegraphics[width=0.55\linewidth]{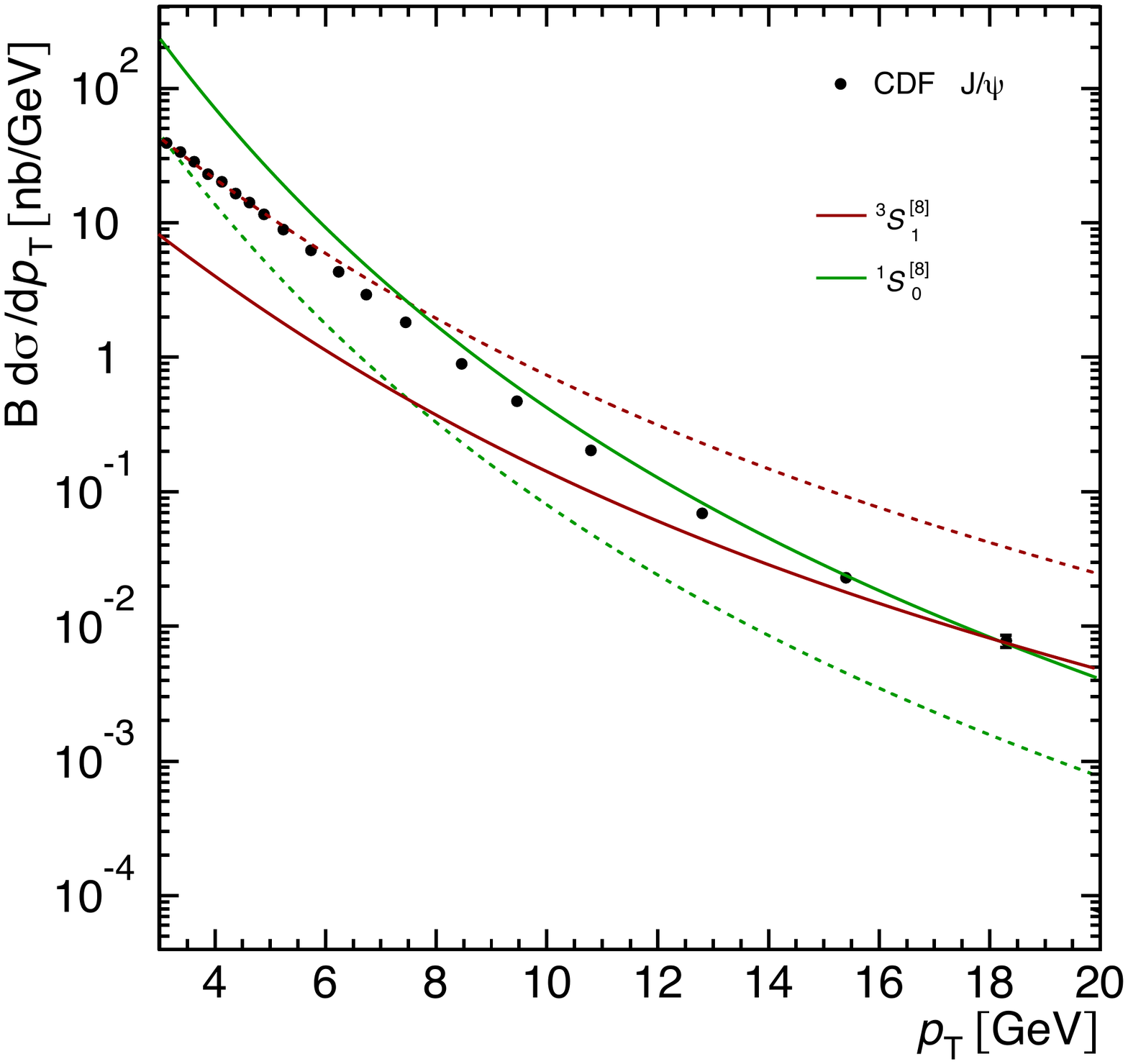}
\caption{$^3S_1^{[8]}$ and $^1S_0^{[8]}$ SDCs of \jpsi\
production~\cite{bib:BKmodel} normalized to 
the first data point (dashed lines) or to
the last data point (solid lines) of the 
CDF measurement~\cite{bib:CDFjpsi2005}.}
\label{fig:lowVShighPTfit}
\end{figure}

One may wonder whether this conclusion is in
contradiction with NRQCD, given that, as seen in the previous
section, current state-of-the-art analyses, based on the fit of only
\pt\ distributions, point to a mixture of the $^1S_0^{[8]}$,
$^3S_1^{[8]}$ and $^3P_J^{[8]}$ channels definitely leading to
transverse polarization~\cite{bib:BKNPB,bib:BKMPLA}.
The answer is that the process mixture resulting from the fit
depends in a dramatic way on the \pt\ range where the fit
is performed. We can have opposite physical indications when the
data are fitted down to the lowest measured \pt\ (the results being
dominated by the more precise low-\pt\ data points) or when we
assume a ``validity domain'' of the theory starting from a higher
\pt\ value. To illustrate this statement, Fig.~\ref{fig:lowVShighPTfit} shows
how the $^3S_1^{[8]}$ and $^1S_0^{[8]}$ SDCs for \jpsi\ production
compare to the data when they are
normalized to the lowest- or highest-\pt\ point.
While the low-\pt\ points are well described by the $^3S_1^{[8]}$
contribution and determine, therefore, a prediction of transverse
polarization if included in the fit, at higher \pt\ the data are
closer in shape to the $^1S_0^{[8]}$ SDC: a fit starting at a \pt\ 
value in the range 10--15~GeV would lead to a dominance of the 
unpolarized $^1S_0^{[8]}$ contribution.
This observation illustrates the crucial importance of performing a
scan of kinematic thresholds to search for a possible domain of
validity of the theory. This procedure will be the subject of
Section~\ref{sec:results_vs_pTmin}. 

Despite often-heard claims to the contrary, a careful look at 
Fig.~\ref{fig:SDCatNLO} reveals that the fitted curve
is a very unsatisfactory description of the measurements, 
given their rather small uncertainties. 
It is usually argued that theoretical uncertainties, not included 
in the fit, can cover the observed discrepancy, reconciling theory 
and data.
However, as shown in Figs.~\ref{fig:SDCatNLO} 
and~\ref{fig:individualPols}, the 
$^3P_J^{[8]}$ octet is the only component that significantly
changes from LO to NLO, in polarization and in the \emph{shape} 
of the \pt\ distribution (changes in normalization are absorbed
in the LDMEs and do not affect the fit quality).
Actually, judging from the difference between the LO and NLO 
calculations, the current theoretical uncertainty in the $^3P_J^{[8]}$ 
term is so large that by considering it in the fit we would introduce an 
excessive freedom, running the risk that this undetermined contribution 
would artificially absorb the data-theory discrepancy: the fit would improve
its ``mathematical quality" at the expense of losing all its physical impact.

We should also mention that, particularly in cases where a
model does not describe faithfully the data, the fit can lead to
meaningless and unstable results. It is helpful, at least as an
initial step --- in our case, the kinematic domain scan --- to reduce
the freedom of the fit to a minimum of essential
parameters, with the aim of obtaining stable and univocal results in
each tested condition. Besides its large uncertainty, the
mathematical peculiarities of the $^3P_J^{[8]}$ SDC, mentioned in
Section~\ref{sec:definitions}, represent a further danger to the
robustness of the fit. Therefore, we will perform our domain scan
considering only the $^3S_1^{[1]}$, $^3S_1^{[8]}$ and $^1S_0^{[8]}$
components.
More than a practical solution, this choice emerges from the
previously discussed data-driven expectation that, within the domain of
validity of the theory, the $^1S_0^{[8]}$ octet must be the dominating
contribution; the $^3P_J^{[8]}$ term, with its unphysical polarization, 
can only represent a relatively small correction. In any case, the impact 
of this initial assumption will be tested a posteriori.

\section{Kinematic domain scan}
\label{sec:results_vs_pTmin}

Our analysis considers a total of 121 data points, measured in pp 
collisions at 7~TeV by three LHC experiments:
ATLAS 
(\upsThreeS\ cross sections~\cite{bib:ATLASYnS2013}), 
CMS 
(\psip~\cite{bib:CMSjpsi2012} and 
\upsThreeS~\cite{bib:CMSYnS2013} cross sections;
and \psip~\cite{bib:CMSpsiPol} and
\upsThreeS~\cite{bib:CMSUpsPol} polarizations)
and LHCb 
(\psip~\cite{bib:LHCbpsip2012} and 
\upsThreeS~\cite{bib:LHCbYnS2012} cross sections).
They correspond to \psip\ data of $p_{\rm T} > 4$~GeV (43 points) and 
to \upsThreeS\ data of $p_{\rm T} > 10$~GeV (78 points), including 
\pt-differential cross sections (99 points) and polarizations
($\lambda_\vartheta$, 22 points). 
We only consider \psip\ and
\upsThreeS\ production because these states are not significantly
affected by feed-down effects and can be treated as being directly produced. 
The description of the production of the remaining \mbox{$S$-wave} quarkonia
contains a considerably higher number of free parameters, the LDMEs
of the $\chi_1$ and $\chi_2$ states,
and will be addressed more 
appropriately when additional constraints from detailed measurements 
of the \mbox{$P$-wave} cross sections and polarizations will become available.

The experiments have provided a thorough account of the dependence
of each cross-section data point on
the polarization hypothesis assumed in the acceptance determination.
It is crucial to take this effect properly into account because it induces a
strong correlation between the cross-section data points and the
actual polarization prediction tested in the fit, thereby correlating
the data points themselves. In our fit procedure, for
each value of the explored parameter space
(i.e., the $^{1}S_0^{[8]}$ and $^{3}S_1^{[8]}$ LDMEs of the
\psip\ and \upsThreeS: four free parameters) we start by calculating the
polarization prediction for each cross-section measurement; then we
use this polarization value to recalculate the cross sections, using an
acceptance correction taken from the tables provided by 
the experiments (with suitable interpolations). We also explicitly treat
the point-to-point correlations induced by the luminosity
uncertainties in the cross-section measurements. To do this, for
each cross-section data set we introduce a nuisance parameter
representing a global rescaling of all points, constrained according to 
the relative luminosity uncertainty.

Concerning the theoretical ingredients, 
we use SDCs (and their longitudinal and
transverse components) calculated for the production of a \QQbar\
pair of $E_0 = 3$~GeV rest energy~\cite{bib:BKmodel}. 
To obtain the shape of the SDC
for the production of a \QQbar\ object of rest energy equal to the 
mass of the considered quarkonium state, $M$, we must rescale the \pt\
variable by $M/E_0$. It can be objected that this is not sufficient,
because the rest energy of the \QQbar\ pair, $E_{Q \overline{Q}}$, is not
necessarily equal to $M$. However, we must also consider what
happens, from the kinematic point of view, in the transition from
the \QQbar\ to the observable quarkonium, because of the emission or absorption
of soft gluons. It can be shown that the average quarkonium
three-momentum $p$ and the \QQbar\ three-momentum $P$ (both in the
laboratory) are related by the approximate expression $\langle p^2
\rangle / P^2 \simeq M^2 / E_{Q \overline{Q}}^2$. The approximation is
excellent (corrections $\leq 2\%$) for $|E_{Q \overline{Q}} - M|$ of the
order of the energy splitting between the radial and orbital angular
momentum excitations of quarkonia. Towards mid-rapidity we can
assume that, on average, $p_{\rm{T}}/p_{\rm{T}}^{Q \overline{Q}} \simeq M
/ E_{Q \overline{Q}}$. Even if, assuming that $E_{Q \overline{Q}}$ were known,
we rescaled the \QQbar\ \pt\ by $E_{Q \overline{Q}}/E_0$ to obtain the
observed quarkonium kinematics, we should then also scale the \pt\
by $M / E_{Q \overline{Q}}$. The net result of the two scalings is
equivalent to one overall scaling by $M/E_0$, which is, therefore,
an as-much-as-possible accurate representation of the
quarkonium production kinematics.

We complement the \pt\ rescaling of the
SDCs with a normalization rescaling exponentially depending on the
quarkonium mass, approximately reflecting the normalization
dependence of the measured $p_{\rm{T}}/M$ distributions
(Fig.~\ref{fig:pToverMscaling}). Since any normalization shift in
the SDC $\mathcal{S}$ is effectively reabsorbed in a rescaling of
the corresponding LDME $\mathcal{L}$ (except for the singlet term,
which gives a negligible contribution), this choice has no influence
on the fit quality nor on the results for the cross sections of the
individual octet processes, $\sigma(A+B \to \QQbar[^{2S+1}L_J^{[8]}]
\to H+X) = \mathcal{S}(A+B\to \QQbar[^{2S+1}L_J^{[8]}] +X) \times
\mathcal{L}( \QQbar[^{2S+1}L_J^{[8]}] \to H)$. Only these cross sections,
denoted by $\sigma(^{2S+1}L_J^{[8]})$ in what follows, can be
directly compared among analyses; the LDMEs fitted from the
\psip\ data, for instance, differ by about a factor of two if we use
the unscaled SDCs.

From a physical point of view, our redefinition of the SDCs
equalizes the meaning of one given LDME among different states: two
states of different mass but same value of $\mathcal{L}(
\QQbar[^{2S+1}L_J^{[8]}] \to H)$ are characterized, with this
convention, by approximately the same probability of the
$\QQbar[^{2S+1}L_J^{[8]}] \to H$ transition; using the 
$E_0 = 3$~GeV convention for both \psip\ and \jpsi, for example, 
the two probabilities would differ by about a factor of two.

As explained in the previous sections,
our central question is
whether it is possible to define a kinematic domain, at sufficiently
high \pt, where current NRQCD calculations give a statistically satisfying
description of the available data. The first step in our analysis
is, therefore, a series of fits performed by selecting only the data
points (for cross sections and polarizations, from all considered
experiments) satisfying the selection $p_{\rm{T}} >
p_{\rm{T}}^{\rm{min}}$, for a progressively changing choice of
$p_{\rm{T}}^{\rm{min}}$.
Following the data-driven motivations given in
Section~\ref{sec:indications}, we consider the LDMEs
$\mathcal{L}(^1S_0^{[8]} \to H)$ and $\mathcal{L}(^3S_1^{[8]} \to
H)$ as the only two parameters of interest, assuming that
$\mathcal{L}(^3P_J^{[8]} \to H)$ is negligible.
For stability reasons, we perform the $p_{\rm{T}}^{\rm{min}}$ scan
without including a modelling of the theoretical uncertainties
(which will be discussed in Section~\ref{sec:results_best}). 
In the absence of theoretical uncertainties, the fits to \psip\ and
\upsThreeS\ data are essentially decorrelated and can be treated 
as two independent procedures.

\begin{figure}[h]
\centering
\includegraphics[width=0.5\linewidth]{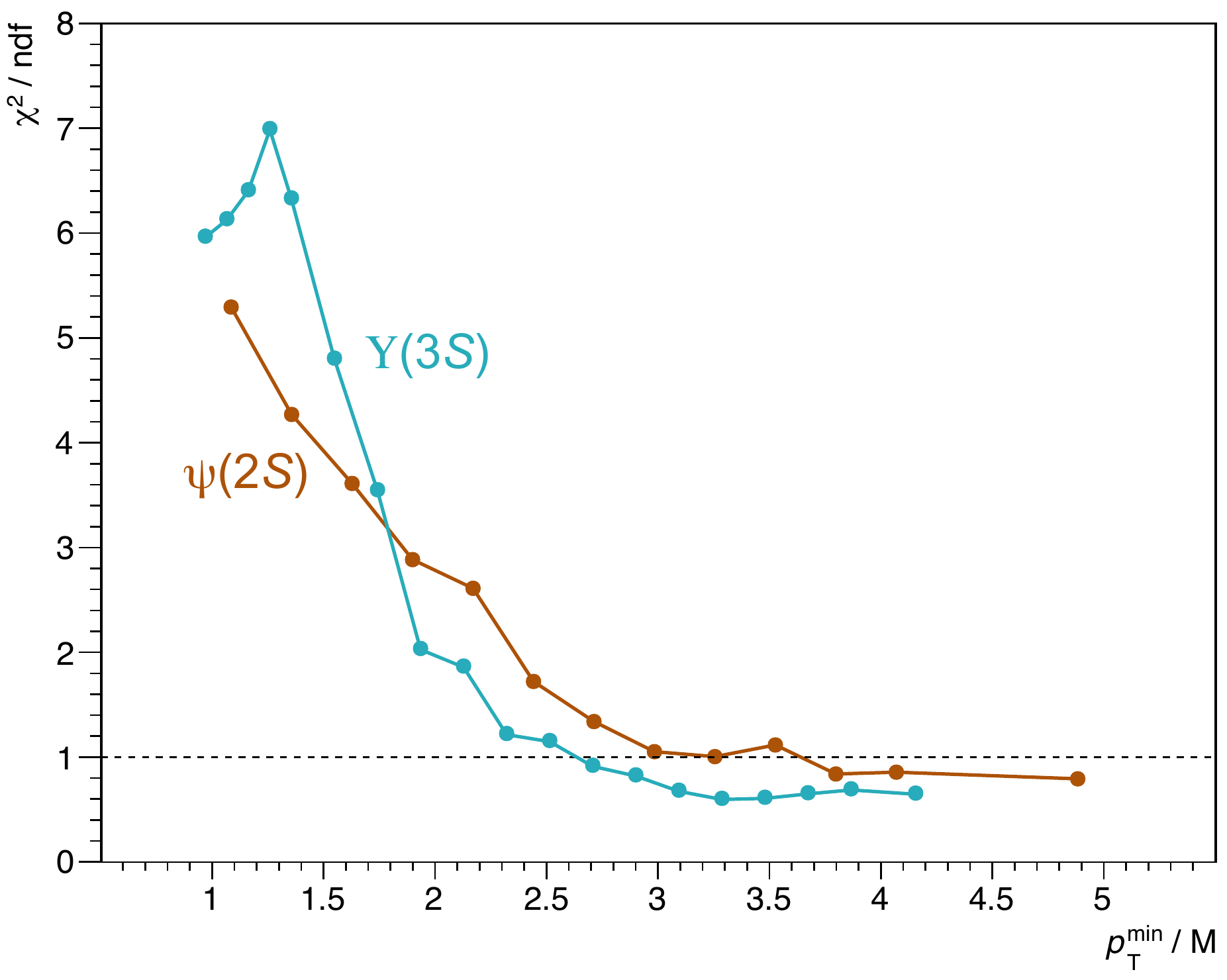}
\caption{Normalised $\chi^2$ of the fits to the \upsThreeS\ and 
\psip\ data, as a function of the mass-scaled \pt\ threshold used 
to select the data, $p_{\rm{T}}^{\rm{min}}/M$.}
\label{fig:scanChi2}
\vglue 3mm
\centering
\includegraphics[width=0.48\linewidth]{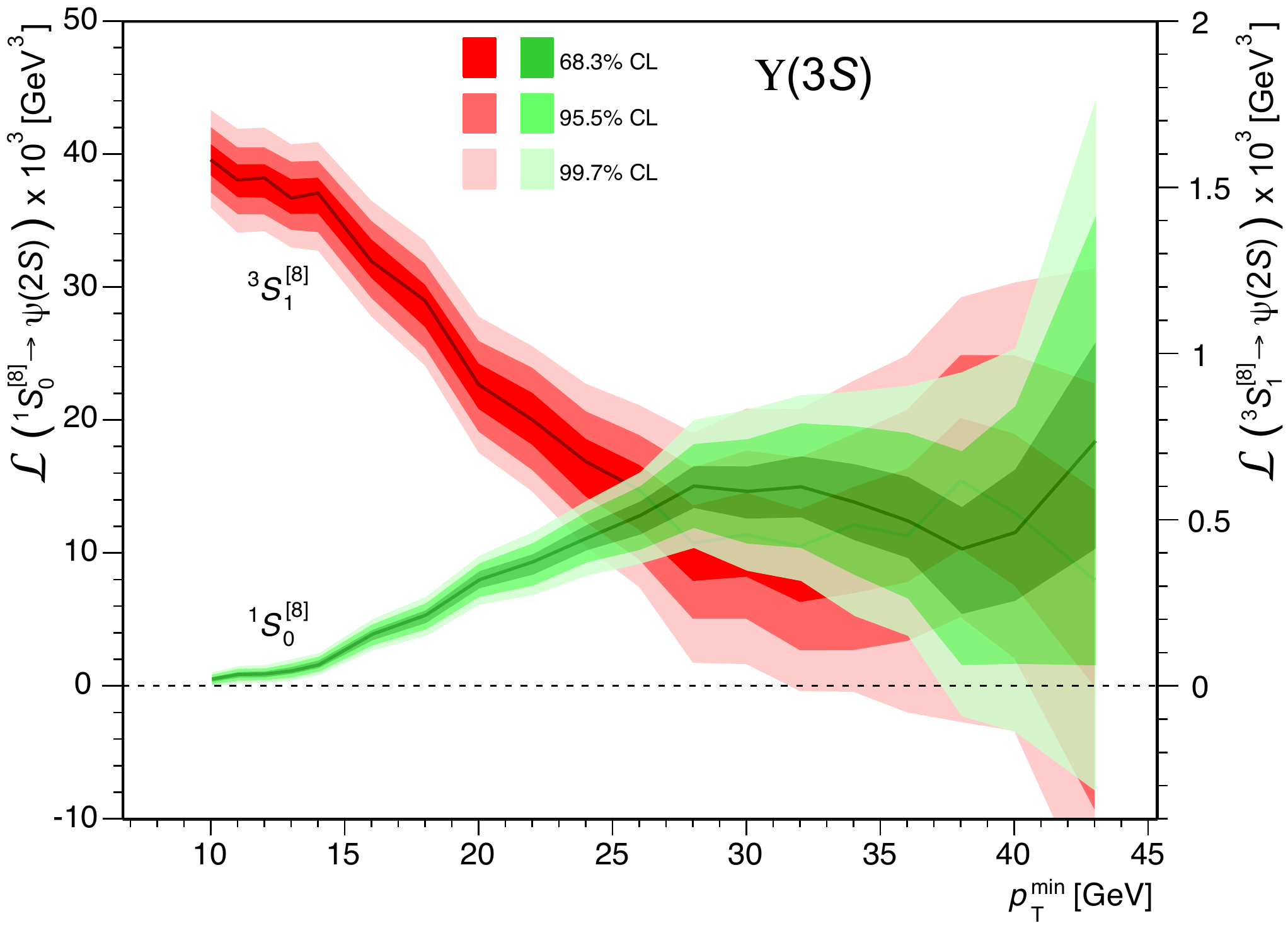}
\includegraphics[width=0.48\linewidth]{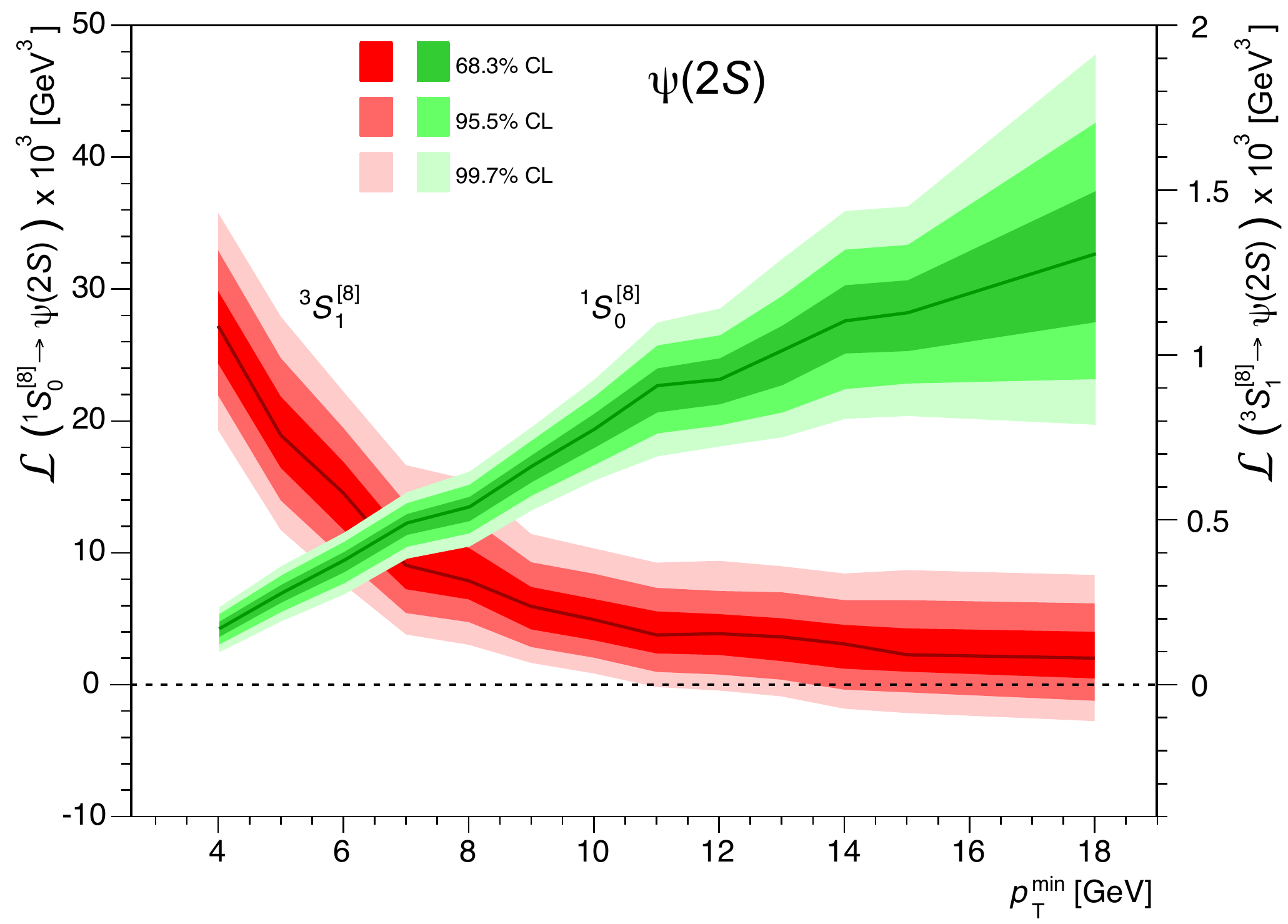}
\caption{Dependence with $p_{\rm{T}}^{\rm{min}}$ of the 
$^{1}S_0^{[8]}$ (green, left y-axes) and $^{3}S_1^{[8]}$ 
(red, right y-axes) LDMEs 
fitted from the \upsThreeS\ (left) and \psip\ (right) data.
See the text for details.}
\label{fig:scanLDMEs}
\end{figure}

Figure~\ref{fig:scanChi2} shows how drastically the quality of the
two fits change with varying $p_{\rm{T}}^{\rm{min}}$, while
Fig.~\ref{fig:scanLDMEs} shows the corresponding behaviour of 
the fitted parameters.

\begin{figure}[ht]
\centering
\includegraphics[width=0.65\linewidth]{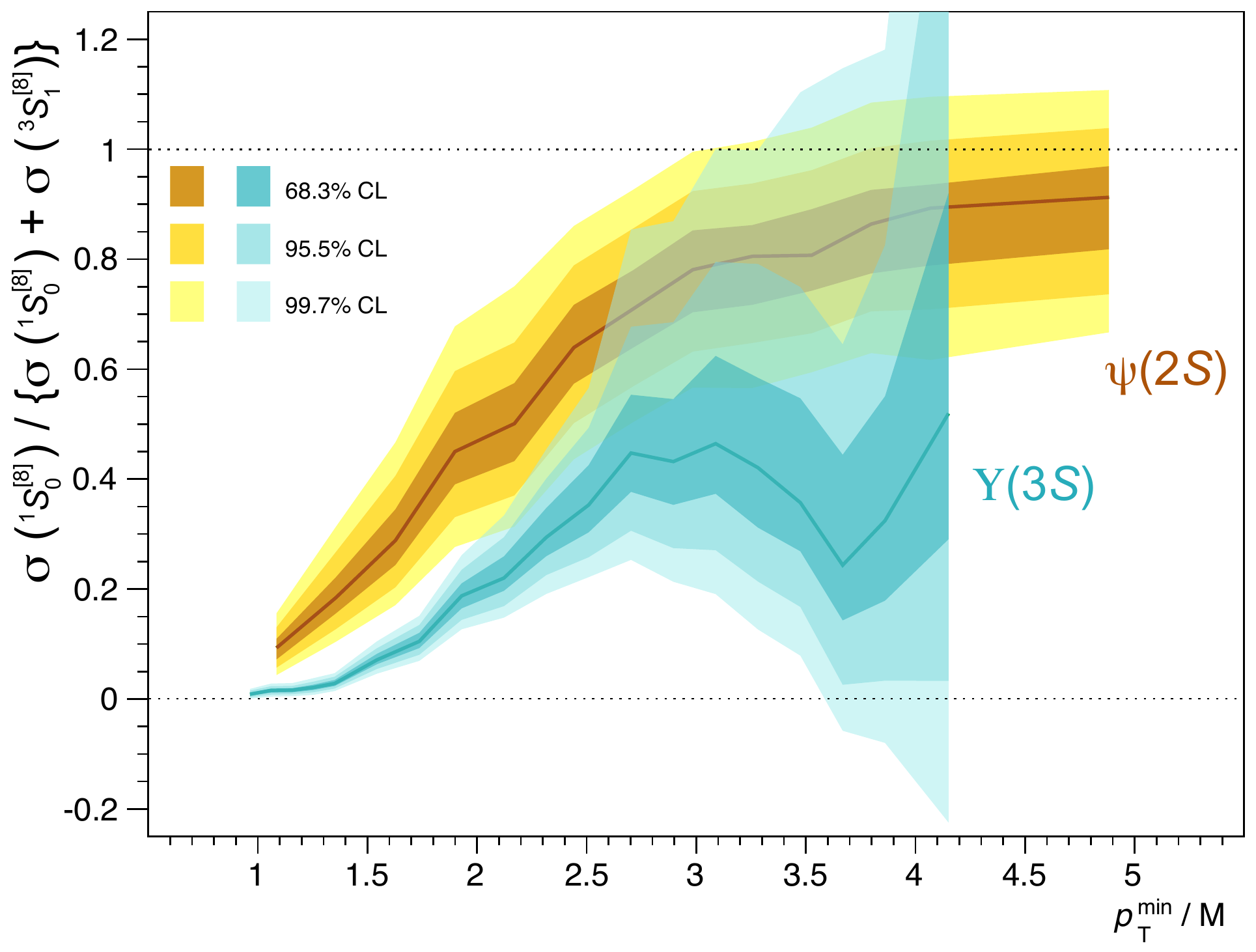}
\caption{Dependence with $p_{\rm{T}}^{\rm{min}} / M$ of the
$\sigma(^{1}S_0^{[8]})$ fraction in the total octet cross section,
calculated for $p_{\rm{T}}/M = 6$.}
\label{fig:scanFractions}
\end{figure}

Let us first consider the \upsThreeS\ case. Above
$p_{\rm{T}}^{\rm{min}}/M \sim 3$ the normalized $\chi^2$ of the fit
stops showing a decreasing trend and reaches a value of order 1 
(the exact value being lower than one simply indicates the presence 
of correlations in the published point-to-point systematic uncertainties
and is not relevant for our studies). 
Correspondingly, the LDMEs
cease to show any systematic trend above 
$p_{\rm{T}}^{\rm{min}} \sim 30$~GeV and start  a 
statistical-like oscillation around
a common value, with obviously increasing uncertainty. This
behaviour is typical
of the stabilization of the
fit results, when tensions between data and theory disappear and
further rejection of data already removes points inside the domain
of the theory. With this criterium, we can consider the \upsThreeS\
results as well described by the theory for \pt\ above $\sim 30$~GeV.
Also in the \psip\ scan we reach a small and rather stable 
normalized $\chi^2$
value, even if the trend of the LDMEs still leaves open the
possibility that a complete stabilization may only happen at
slightly higher values of $p_{\rm{T}}^{\rm{min}}$. 

While future data, extending with
better precision towards higher \pt, are needed for a conclusive
statement, we do not expect a significant
change of the physical conclusions, as can be appreciated from
Fig.~\ref{fig:scanFractions}. The relative importance of
the $\sigma(^{1}S_0^{[8]})$ colour-octet cross section with respect
to the total contribution of colour-octet processes, calculated
at an arbitrary reference $p_{\rm{T}}/M = 6$ and mid-rapidity, 
saturates close to unity in the \psip\ case, clearly indicating 
that the $^{1}S_0^{[8]}$ octet state dominates \psip\ production,
whereas the \upsThreeS\ trend points to a more
democratical share between the $^{1}S_0^{[8]}$ and $^{3}S_1^{[8]}$
contributions, at such high \pt\ values.

Figure~\ref{fig:scanFractions} also shows, for both quarkonia, that
a fit performed by undiscriminatingly including
all available data down to the lowest \pt\ would lead, with high
significance, to the opposite physical conclusion: that
$^{1}S_0^{[8]}$ production is negligible and the \mbox{$S$-wave} cross
sections are dominated by the (transversely polarized)
$^{3}S_1^{[8]}$ contribution. As anticipated in
Section~\ref{sec:indications}, this conclusion,
``traditionally'' presented as a prediction of NRQCD, is in reality 
a result completely determined by the use of data not
belonging to the domain of validity of the theory calculations.

We conclude this section by clarifying that our considerations
would not be modified by the inclusion of photoproduction data, 
given that all such measurements are presently restricted to the 
low-\pt\ region, excluded by our study. Therefore, the hypothesis 
that the LDMEs are universal cannot be tested until precise
photoproduction data will become available at high \pt.

\section{Results and predictions}
\label{sec:results_best}

Given the results shown in the previous section, we continue our
analysis only using the 44 data points (30 cross sections and 14
polarizations) that belong to the kinematic domain 
$p_{\rm T}^{\Upsilon(3S)} > 30$~GeV and 
$p_{\rm T}^{\psi(2S)} > 12$~GeV. These numerical values are
clearly affected by some degree of arbitrariness and might have to
be adjusted, at least in the \psip\ case, when more precise
high-\pt\ data will become available.

We start by addressing our data-driven assumption that we can
neglect the $^{3}P_J^{[8]}$ contributions in the description of
\mbox{$S$-wave} quarkonium production. Given the very good quality of the
fits performed with $\mathcal{L}(^{3}P_J^{[8]})=0$, the current
\upsThreeS\ and \psip\ are far from indicating the necessity of a
non-vanishing \mbox{$P$-wave} octet component. 
Despite the caveats exposed in Sections~\ref{sec:definitions} 
and~\ref{sec:indications} (in particular, we must be very critical 
regarding fits including this octet component, given its overwhelming 
theoretical uncertainty), we have repeated the fit
with the additional free parameter $\mathcal{L}(^{3}P_J^{[8]})$. In
the \psip\ case, the central values of the fit results do not
change, the fractional contributions to the octet cross
sections at $p_{\rm{T}}/M=6$ being $(80 \pm 8)\%$
($^{1}S_0^{[8]}$), $(20 \pm 20)\%$ ($^{3}S_1^{[8]}$) and $(0 \pm
20)\%$ ($^{3}P_J^{[8]}$), fully consistent with our hypothesis and
with the result shown in Fig.~\ref{fig:scanFractions}. The
large $^{3}S_1^{[8]}$ and $^{3}P_J^{[8]}$ uncertainties are strongly
anti-correlated. The \upsThreeS\ fit becomes strongly
under-constrained, still favouring, nevertheless, the $^{1}S_0^{[8]}$
octet-cross-section component (at $80^{+70}_{-30}\%$).

For a realistic evaluation of the LDMEs,
we have included theoretical uncertainties in the fit procedure. 
For the $^{1}S_0^{[8]}$ and
$^{3}S_1^{[8]}$ SDCs and polarizations, we assume as uncertainty
(corresponding to a $\pm 1\sigma$ variation) the magnitude of the
difference between NLO and LO calculations (shown in
Section~\ref{sec:definitions}). In the case of the colour-singlet cross
section and polarization, we keep the NLO calculation as central
model but define the uncertainty as the difference with respect
to the partial NNLO calculation of Ref.~\cite{bib:CSpartialNNLO},
except for the $- 1\sigma$ uncertainty in the cross section case, taken to be
the LO model, to constrain the cross section to positive values. The
nuisance parameters describing these allowed variations of the
short-distance ingredients are kept common to the \upsThreeS\ and \psip\
in one global fit, accounting for the correlation that the theoretical 
uncertainties induce between them.

\begin{figure}[p]
\centering
\includegraphics[width=0.9\textwidth]{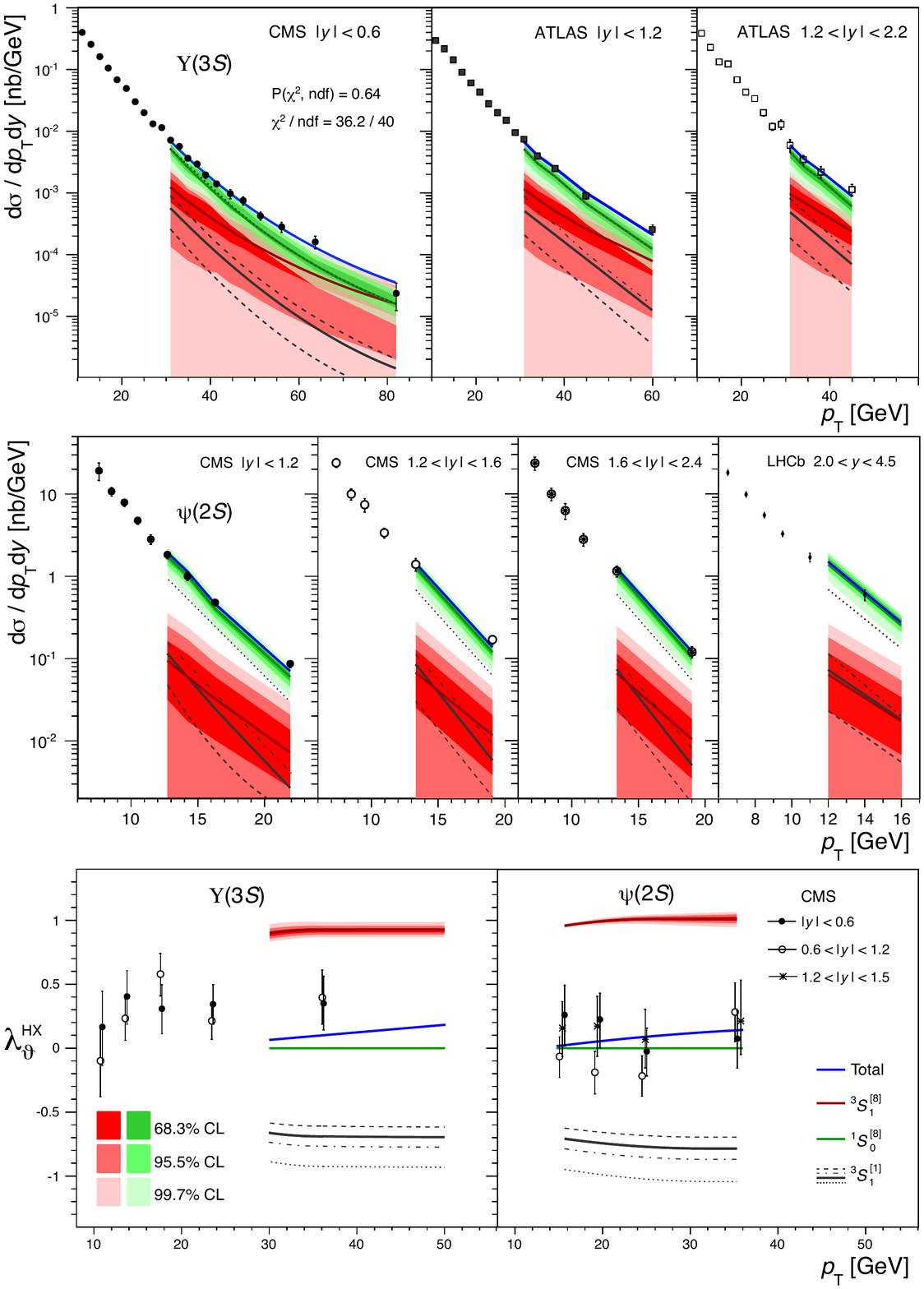}
\caption{Data points used in the global fit and resulting curves for
the total and individual terms. See text for details.}
\label{fig:globalfit}
\end{figure}

Figure~\ref{fig:globalfit} shows the fitted data and the
best-fit curves for the cross sections and polarizations, including
the individual colour-singlet and colour-octet contributions.
Uncertainty bands are also shown for the
$^{1}S_0^{[8]}$ and $^{3}S_1^{[8]}$ cross section and polarization
components,
while the colour-singlet contribution is represented by the 
LO (dashed), NLO (dot-dashed) and partial NNLO (dotted)
calculations, together with the corresponding best-fit curve (solid),
which is lower than the NLO calculation.


\begin{figure}[t!]
\centering
\includegraphics[width=0.55\linewidth]{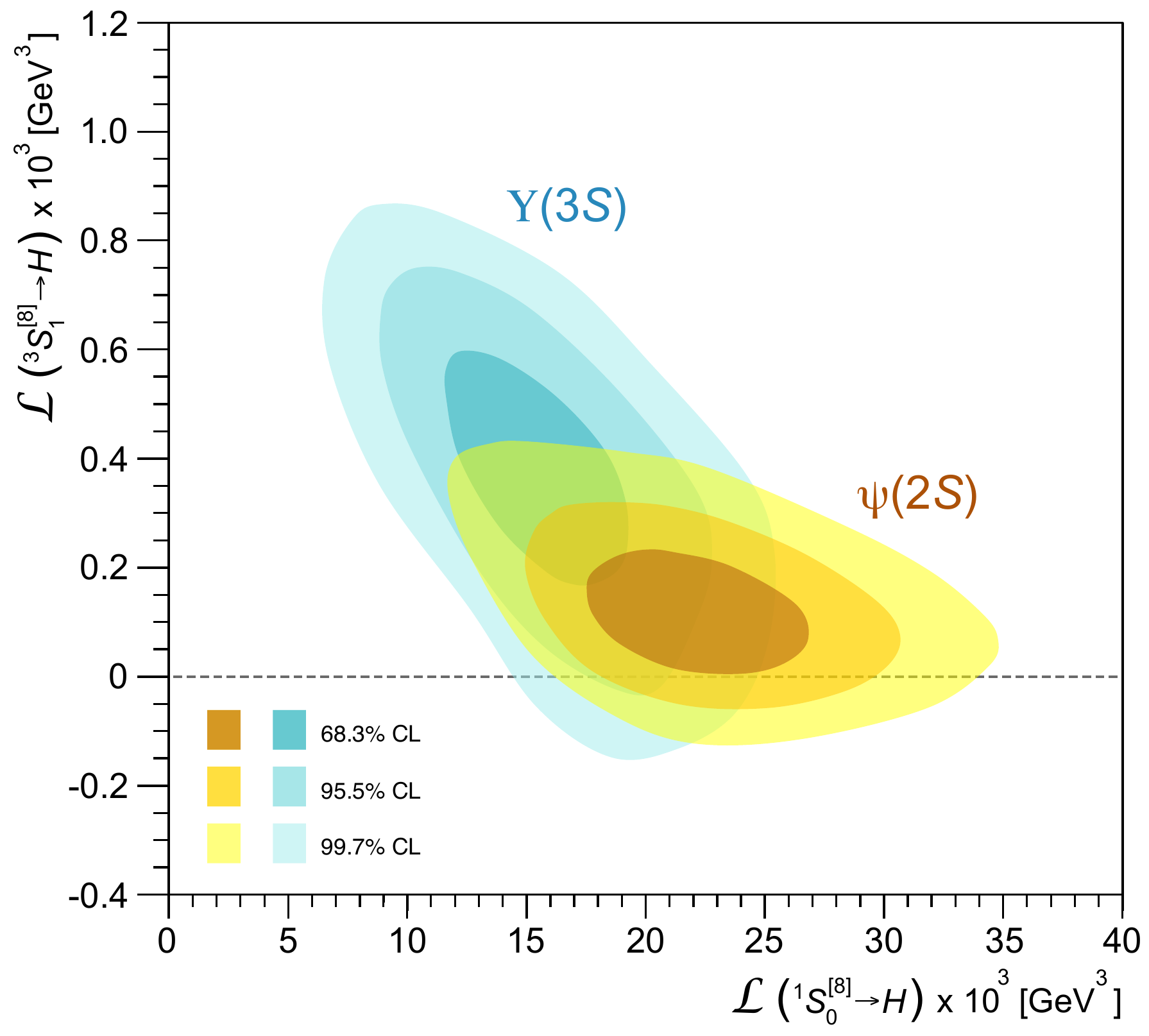}
\caption{Probability densities of the fitted $^{1}S_0^{[8]}$ and 
$^{3}S_1^{[8]}$ LDMEs, for the \psip\ and \upsThreeS, represented 
by the 68.3\%, 95.5\% and 99.7\% confidence level contours.}
\label{fig:2Dcontours}
\vglue 3mm
\centering
\includegraphics[width=0.55\linewidth]{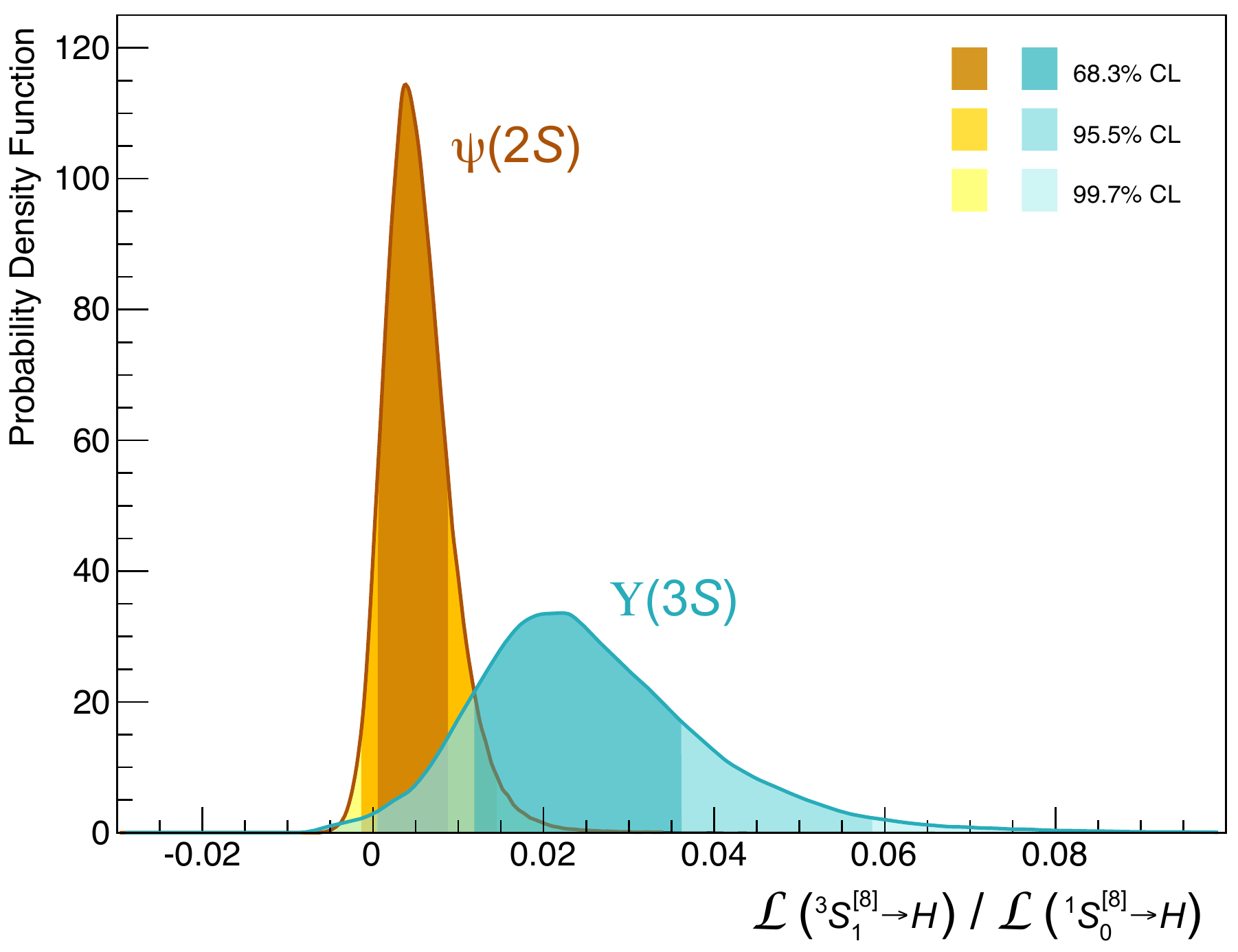}
\caption{\psip\ and \upsThreeS\ probability densities of the 
$^{3}S_1^{[8]} / ^{1}S_0^{[8]}$ ratio of LDMEs.}
\label{fig:LDMEratio}
\end{figure}

Figure~\ref{fig:2Dcontours} shows the \psip\ and \upsThreeS\
probability densities of the fitted $^{1}S_0^{[8]}$ and $^{3}S_1^{[8]}$ 
LDMEs, in the form of two-dimensional contours, while 
Fig.~\ref{fig:LDMEratio} shows the corresponding distributions of the
$^{3}S_1^{[8]} / ^{1}S_0^{[8]}$ LDME ratio. 
Remarkably, the magnitudes of the two matrix elements are very 
different, in contradiction with usual expectations, as discussed in
the next section.

\begin{figure}[h]
\centering
\includegraphics[width=0.48\linewidth]{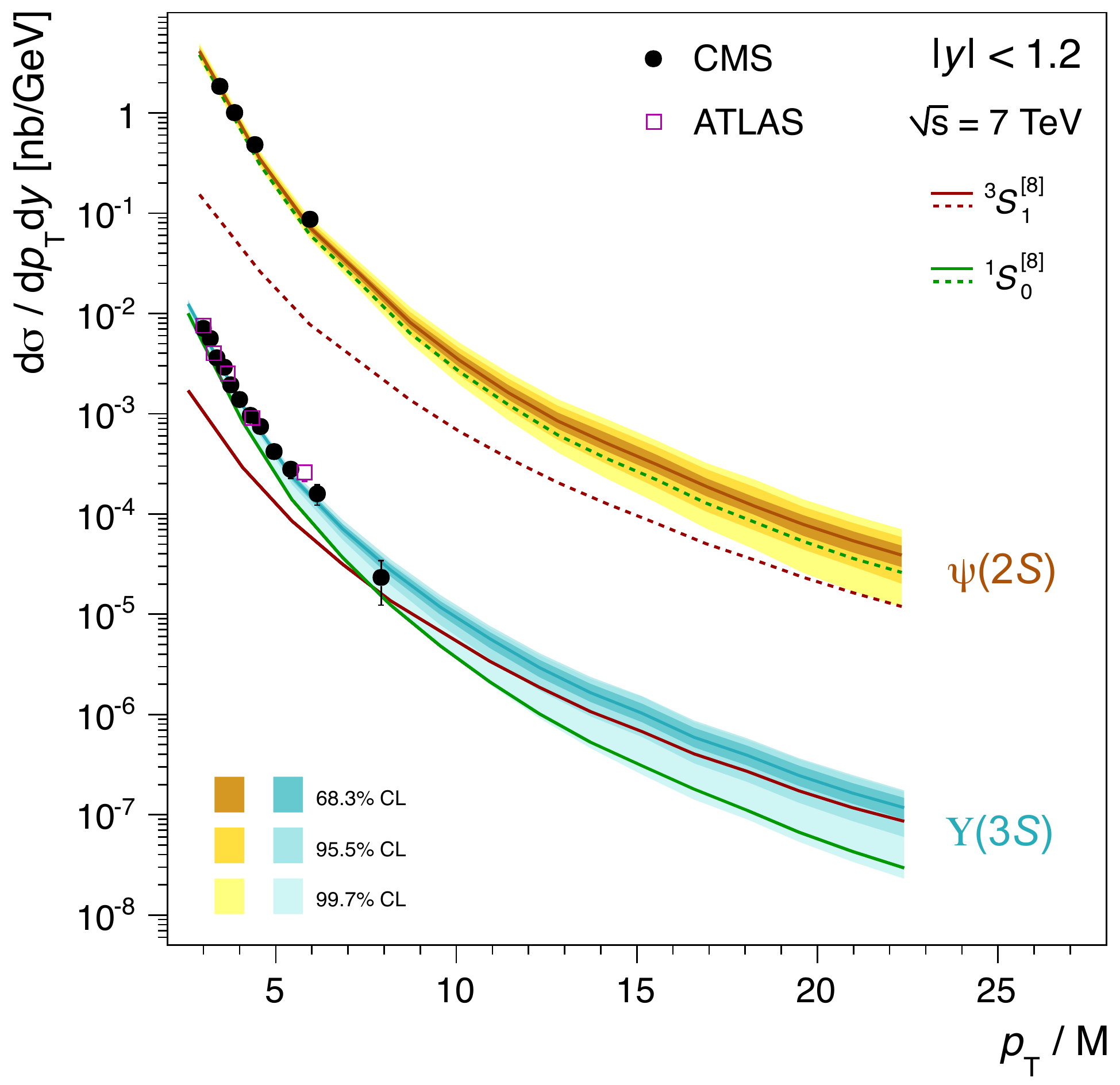}
\includegraphics[width=0.48\linewidth]{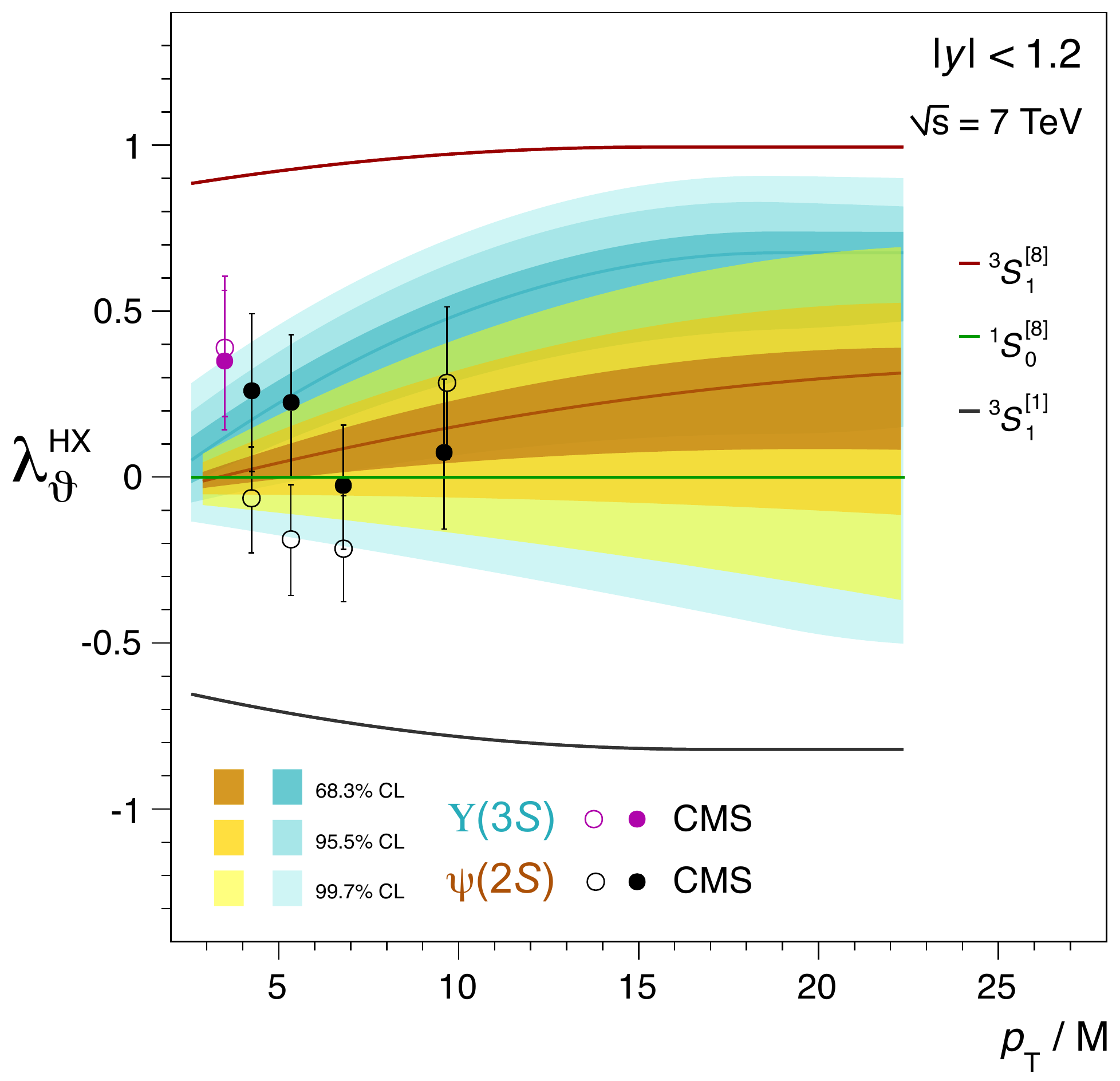}
\caption{\psip\ and \upsThreeS\ mass-scaled \pt-differential cross sections 
(left) and polarizations (right) extrapolated to much higher \pt\ values than
the ranges covered by the fitted data, also shown on the figures.}
\label{fig:prediction}
\end{figure}

Overall, in both the \psip\ and \upsThreeS\ cases, the $^{1}S_0^{[8]}$ 
octet is the dominating production channel. However, at very high \pt\
the $^{3}S_1^{[8]}$ contribution seems to start prevailing in the
\upsThreeS\ case, as can be observed in the top left panel of 
Fig.~\ref{fig:globalfit}, indicating that very-high-\pt\ \upsThreeS\ 
mesons might be produced with a strong transverse polarization.
This is clearly illustrated in
Fig.~\ref{fig:prediction}, which shows the cross sections and polarizations 
corresponding to the fitted LDMEs, extrapolated to \pt\ values well 
beyond the ranges probed by the existing measurements.
These predictions were calculated using the two-dimensional 
probability-density function for the fit parameters, thereby taking 
into account parameter correlations and the modelled theoretical 
uncertainties, besides the experimental ones.

To put our results in the context of the existing literature, we
stress that this is the first analysis specifically dedicated to the
strategy for the comparison of existing theory calculations to
measurements, with a data-driven attitude and a focus on the
treatment of the experimental results. Previous ``global-fit" 
analyses were, instead, byproducts of studies centred in the 
calculation of the short-distance ingredients of NRQCD. 
In fact, remarkable efforts by a few groups triggered a great 
progress on this front over the last years, leading to full NLO 
cross-section and
polarization calculations for different collision systems, energies
and kinematic domains, for all relevant colour-singlet and
colour-octet processes, including \mbox{$P$-waves}.
The comparisons with data, however, have not followed
detailed strategies and rigorous reproducibility criteria, so that it 
is often difficult to
appreciate the consequences and prospects of the different fit
approaches. It is sometimes impossible to understand the reasons of
the differences in the fit results, which in some cases are very significant,
giving the wrong overall impression that the NRQCD framework is
either very unstable with respect to variations in the inputs or
that it can at most give order-of-magnitude evaluations of 
cross sections and qualitative estimates of polarizations.

Some analyses include cross-section measurements down to 
$p_{\rm{T}} = 3$~GeV, others apply a fixed threshold $p_{\rm{T}} > 7$~GeV. 
With different data sets from analysis to analysis, it is
difficult to quantify exactly how the choice affects the results
and the quality of the theory-data agreement. In fact,
the quantification of the agreement only addresses the cross sections
or is not even reported. Moreover, since the polarization uncertainty 
correlations and luminosity uncertainties are never mentioned, one has
to assume that they are neglected or assumed to be uncorrelated among 
different kinematic intervals, a choice that introduces an artificial freedom 
in the predicted shapes.

We take as examples the three recent analyses of prompt charmonium
production reported in Refs.~\cite{bib:BKNPB,bib:BKMPLA} (A1), 
\cite{bib:Chaofit} (A2) and~\cite{bib:Gaofit} (A3).
A1 considers a large amount of \jpsi\ data from Tevatron, LHC, RHIC
and photo-production, using only \pt\ distributions as constraints
and neglecting the feed-down from $\chi_c$ decays.
A2 only considers \jpsi\ data from CDF, for $p_{\rm{T}} > 7$~GeV,
including the polarization as constraint and neglecting the $\chi_c$
feed-down.
A3 uses CDF and LHCb data for \jpsi, \psip\ and $\chi_c$, with
$p_{\rm{T}} > 7$~GeV, excluding polarizations and including the
modelling of the feed-down for the \jpsi.
The outcomes of A1 and A3 are, despite the very different strategies,
substantially similar: a strong transverse polarization is predicted
for directly produced $S$-wave charmonium in the \pt\ range covered 
by the CMS measurements.
A2 reproduces the unpolarized scenario by allowing a mutual
cancellation of the transverse polarizations of the $^3S_1^{[8]}$
and $^3P_J^{[8]}$ cross-section terms, which are found to have
opposite signs. However, the study suggests that no unique scenario
can describe at the same time the CDF \pt\ distributions 
($7< p_{\rm{T}} < 20$~GeV) and the LHC ones ($7 < p_{\rm{T}} < 70$~GeV).
The latter are shown to lie entirely, with their error bars, between
two curves, corresponding to the octet cross section term containing
either 0\% or 100\% contribution of $^1S_0^{[8]}$, which exclude the CDF best-fit
result. It is concluded that the current LHC measurements lack constraining 
power on the parameter space, especially on the $^1S_0^{[8]}$ LDME.
No quantification of the goodness of agreement is reported for any of
the considered scenarios.

\section{Discussion on the observed LDME hierarchies}
\label{sec:discussion}

The results of our study point to the existence of a much stronger
hierarchy of LDMEs than the one predicted by the usual
power-counting scheme of NRQCD, based on elegant and very general
considerations on the formal structure of the NRQCD Lagrangian and
operators. In NRQCD the three octet contributions $^1S_0^{[8]}$,
$^3S_1^{[8]}$ and $^3P_J^{[8]}$ are expected to scale in the same
way for small values of the heavy-quark velocity $v$ and,
therefore, to have similar magnitudes. Two basic considerations
compete in the determination of this result. On one hand,
independently of the observed particle (\mbox{$S$-wave} or \mbox{$P$-wave}
quarkonium), transitions from octet (or singlet) states with
non-zero orbital angular momentum are suppressed, reflecting the
fact that the perturbative \QQbar\ state must be produced at
short-distance and small relative momentum. In particular,
transitions from pre-resonance \mbox{$P$-wave} octet (and singlet) states are
suppressed by a factor $v^2$ (coming from two additional spatial
derivatives in the structure of the respective operators). On the
other hand, the probability of soft-gluon emission depends on the
process. The $^3P_J^{[8]} \to \psi / \Upsilon + g$ process is a
chromoelectric transition ($\Delta L = \pm 1$, $\Delta S = 0$),
while $^1S_0^{[8]} \to \psi / \Upsilon + g$ is a chromomagnetic
transition ($\Delta L = 0$, $\Delta S = \pm 1$): their probabilities
scale, respectively, like $v^2$ and $v^4$. The $^3S_1^{[8]} \to
\psi / \Upsilon + gg$ process is predominantly a double-chromoelectric
transition, its probability scaling like $v^4$.
These two considerations alone lead to the prediction that the three
processes should have comparable probabilities (all scaling like $v^4$).

In order to understand why, instead, the data indicate the
hierarchy ${^3P_J^{[8]}} \ll {^3S_1^{[8]}} \ll {^1S_0^{[8]}}$, these rules
must apparently be integrated with further conjectures on the
mechanism of quarkonium formation. For example, the dependence of
the interaction potential on the colour state of the quark-antiquark
pair may play a role. With colour neutralization, the short-distance
potential changes from weakly repulsive, $V_8 \geq 0$, to
attractive, $V_1 < 0$. Therefore, in the octet-to-singlet transition
the \QQbar\ pair undergoes a significant decrease in potential
energy, $\Delta V(8 \to 1) \simeq V_1 \simeq -T$, of the order of
the kinetic energy $T$ of the bound state, i.e., of the energy
splitting between radial and orbital angular momentum excitations of
the quarkonium, $\sim 0.4$--0.6~GeV (very similar for charmonium and
bottomonium). Transitions in which the \QQbar\ kinetic energy
decreases ($\Delta T<0$) should therefore be disfavoured, because
they require that the emitted soft gluons have 
comparatively high energy: $E_g = |\Delta V| - \Delta T$. In
particular, the $^3P_J^{[8]} \to {\rm J}/\psi \, [\Upsilon(1S)]$ transition,
with $\Delta T \sim m({\rm J}/\psi \, [\Upsilon(1S)]) - m(\chi) \sim -0.4$~GeV,
should be suppressed, while the transition from $^1S_0^{[8]}$, with
$\Delta T \sim  m({\rm J}/\psi \, [\Upsilon(1S)]) - m(\eta_c) \sim +0.1$~GeV,
would be the least subjected to the energy requirement on the gluon
radiation. This sort of threshold effect may explain why \psip\
and \upsThreeS\ productions are dominated by the $^1S_0^{[8]}$ and
$^3S_1^{[8]}$ octets. 

Another fact worth of attention
is that the measured $^3S_1^{[8]}$ suppression with respect to
$^1S_0^{[8]}$ is not as strong for the \upsThreeS\ as
for the \psip. This may possibly reflect the fact that the $b$
quark in a bottomonium state, having larger average momentum than
the $c$ quark in a charmonium state, can emit higher-energy gluons.
Clearly, this is only one possible conjecture. Alternative
velocity-scaling schemes~\cite{bib:Schuler,bib:Fleming} also go in
the direction of a better qualitative description of the measured
LDME hierarchy, by reducing or eliminating the relative suppression
of the chromomagnetic octet-to-singlet transition with respect to
the chromoelectric one, therefore favouring the single-emission
transition $^1S_0^{[8]}$ over the double-emission transition
$^3S_1^{[8]}$.
Incidentally, the different quality of the interaction potential for
singlet and octet quark-antiquark pairs may also have a role in the
observed dominance of octet processes over the singlet ones, given
that the expansion of the initial ``point-like'' \QQbar\ towards
bound-state sizes is energetically favoured when the short-distance
potential is repulsive rather than attractive.

These reasonings do not pretend to represent univocal
explanations of the measured effects. They should be considered as 
illustrations of how the observation of definite scaling hierarchies
for the LDMEs as a function of quarkonium mass, binding energy and
quark flavour can have strong implications concerning the
long-distance processes at play. Clarifying such hierarchies is one of 
the most stimulating reasons justifying accurate quarkonium production 
measurements at high-\pt, to be made at the LHC, so that we can pave 
the way towards a clear-cut understanding of bound-state formation in QCD.

Concerning \mbox{$P$-wave} quarkonium production, the double chromoelectric
transition $^3P_J^{[8]} \to \chi + gg$ is disfavoured with respect
to the single $^3S_1^{[8]} \to \chi + g$ one, besides being
suppressed because of the higher angular momentum of the
colour-octet state. Also for \mbox{$P$-wave} quarkonium production we expect,
therefore, that the \mbox{$P$-wave} octet contribution is negligible. Among
the remaining ones, the single chromoelectric transition
$^3S_1^{[8]} \to \chi + g$ should be favoured with respect to the
double chromomagnetic+chromoelectric $^1S_0^{[8]} \to \chi + gg$
transition, but the relative importance of the two may be influenced
by the $\Delta T>0$ enhancement of the latter and, a priori, both should be
taken into consideration. 

The possibility of a
non-negligible role of the $^1S_0^{[8]}$ contribution in $\chi$
production, in analogy with $\psi$ and $\Upsilon$ production, is suggested
by the two experimental facts discussed in Section~\ref{sec:indications}: 
1)~the approximate universality of the $p_{\rm T}/M$ scaling of 
\mbox{$S$-wave} quarkonium cross sections, indicating that states 
with a significant $\chi$ feed-down behave similarly to the others;
2)~the absence of a clear polarization pattern differentiating directly 
produced states from those affected by a large $\chi$ feed-down. 
Future $\chi$ polarization measurements will be crucial to distinguish 
between the $^1S_0^{[8]}$ and $^3S_1^{[8]}$ contributions, respectively
characterized by lack of polarization ($\chi$ from $^1S_0^{[8]}$) or by 
moderate transverse polarizations (high-\pt\ $\chi_1$ and $\chi_2$ from 
$^3S_1^{[8]}$ have $\lambda_\vartheta = +1/5$ and $+21/73$, respectively, 
in the centre-of-mass helicity frame\,\footnote{Values calculated following
the method of Ref.~\cite{bib:chiPol}, applied to the decay chain
$^3S_1^{[8]} \to \chi_{1,2} +g$, $\chi_{1,2} \to \psi/\Upsilon +
\gamma$, where $^3S_1^{[8]}$ has angular momentum projection $J_z =
+1$ or $-1$ (transverse polarization), and assuming electric-dipole
gluon and photon radiations.}).

\section{Summary and conclusions}
\label{sec:summary}

Non-relativistic QCD, a rigorous and consistent effective theory
based on QCD, should provide an accurate description of heavy
quarkonium production. However, the efforts to validate NRQCD as a
working framework have brought to light serious and persistent
mismatches between data and calculations, especially concerning
polarization. Recent CMS measurements of the polarizations of
(directly produced) \psip\ and \upsThreeS\ have seemingly removed
any residual ambiguity in this evidence.

We have addressed the ``quarkonium production puzzle'' through a
deep reconsideration of the strategy for theory-data comparison.
While the polarization data are traditionally excluded from global
NRQCD analyses of quarkonium production (and used only as a
posteriori verifications of the predictions), we argue that they are
actually the most stringent and straightforward constraints in
discriminating the underlying fundamental processes and we move them
from the periphery to the centre of the study.

In fact, the measured unpolarized scenario points to a
straightforward Occam-razor interpretation: the different
colour-octet contributions to the \mbox{$S$-wave} quarkonium yield follow a
magnitude hierarchy reflecting their degree of polarization. The
unpolarized $^1S_0^{[8]}$ channel should dominate, while the
$^3P_J^{[8]}$ one, with a polarization more transverse than
what is physically allowed, should at most be a tiny correction. A
small $^3S_1^{[8]}$ contribution, characterized by a fully transverse
(but physical) polarization, would be sufficient to explain the
possible tendency of the measured polarizations towards slightly
transverse values at higher \pt.

The data show another interesting pattern: the differential cross sections 
of seven quarkonium states are compatible with a common $p_{\rm{T}}/M$ 
scaling, at least for $p_{\rm{T}}/M > 3$.
Given that these quarkonia include two essentially pure \mbox{$S$-wave} 
states (\psip\ and \upsThreeS), three \mbox{$S$-wave} states affected by a 
significant feed-down from \mbox{$P$-wave} states (\jpsi, \upsOneS\ and 
\upsTwoS) and two \mbox{$P$-wave} states (\chicOne\ and \chicTwo),
their very similar behaviour suggests that quarkonium production is the 
result of a simple mixture of processes, stable with varying mass and 
quantum numbers. This observation clearly favours a scenario where one
single process dominates and, together with the polarization argument,
makes it even less reasonable to consider
that the unpolarized measurements could be the result of a 
delicate cancellation of strongly polarized processes.
Furthermore, given that both the $^3P_J^{[8]}$ and $^3S_1^{[8]}$ octets
are transversely polarized, their mutual cancellation implies that one of
them needs to contribute with a negative cross section.

These data-driven considerations guide us in our global fit of LHC
measurements of \psip\ and \upsThreeS\ cross sections and
polarizations. Having a prior expectation of what a reasonable
result will be helps us avoiding the pitfalls of ill-posed,
under-constrained or unstable fits.
By excluding polarization data from the fits, previous analyses have
effectively chosen to restrict the safe domain of the theory to the
description of the unpolarized cross-section observables. We propose
a different definition of field of validity, including polarization
observables as crucial players while possibly excluding the lowest-\pt\
data, knowing that fixed-perturbative-order factorization
calculations are supposed to work only at sufficiently high \pt. 
The systematic search for the domain of validity of the theory through a 
scan of the kinematic phase space is a crucial step in our analysis.

For the first time in this kind of studies, we perform a
rigorous treatment of correlated experimental uncertainties,
including the dependence of experimental acceptances on the
polarizations. Once a candidate domain of validity is defined, we
also include in the fit a modelling of the theoretical uncertainties,
so that they are reflected in the output parameters. This
effectively introduces a partial correlation between the \psip\ and
\upsThreeS\ systems: charmonium and bottomonium fits become one
global quarkonium fit.

Bringing the polarization data to the centre of the stage and decreasing 
the (statistically strongest) weight of the low-\pt\ data is a ``Copernican 
revolution'' that seems to provide a straightforward solution to the
puzzle: the cross sections and the polarizations are both perfectly fitted 
by the theory in a domain approximately defined by the selection cut
$p_{\rm{T}}/M > 3$.
Confirming our initial expectation, no \mbox{$P$-wave} component is 
needed to describe the data. We also find that the data favour a
colour-singlet component smaller than the NLO calculation and even
ten times smaller than the partial NNLO calculation. 

These facts, together with the further hierarchy
$\mathcal{L}(^3S_1^{[8]})  \ll \mathcal{L}(^1S_0^{[8]})$, are physically
intriguing and are to be interpreted as strong indications for the
understanding of the mechanisms of bound-state formation 
(an example of such interpretations being presented in
Section~\ref{sec:discussion}).
%
Furthermore, finding that the $^3P_J^{[8]}$ octet term gives
a negligible contribution to quarkonium production is also 
extremely interesting from another perspective.
In fact, contrary to what happens in the \mbox{$P$-wave} octet case,
the SDCs of the dominant \mbox{$S$-wave} octet components have a
very stable shape from LO to NLO, indicating that, at the present status of the 
perturbative calculations, the theoretical uncertainties in the framework are
relatively small. This points to a great potential of NRQCD as a
precision instrument to address and isolate the intriguing aspect of
the process, the formation of the bound state, as described by the
non-perturbative LDMEs. If these observations are confirmed by future
data, the LHC measurements will provide precise determinations of the
LDMEs of all quarkonium states in a consistent framework. On the
other hand, we must call attention to the fact that the existing
photo-production data belong to the kinematic domain that our study
has excluded. Therefore, a test of the universality of the LDMEs
must wait for precise high-\pt\ measurements in processes
different from direct production in pp collisions.

We must also mention that, while the \pt\ distributions for
$p_{\rm{T}}/M < 3$ cannot be described at NLO using the same process
mixture implied by higher-\pt\ data, they are, nevertheless, still
compatible with the zero-polarization pattern, smoothly
continuing the high-\pt\ trend (see
Fig.~\ref{fig:polMeasurements}). In other words, there is no
indication from data alone of a change in production mechanism from
high to low \pt. The implied dominance of quarkonium production via
an intermediate isotropic wave function (presumably $^1S_0^{[8]}$)
finds its simplest explanation in one of the crucial aspects of the
factorization concept: the quantum numbers of the produced \QQbar\
change during the bound-state formation, making it possible that,
for example, a $J=1$ quarkonium exhibits a distinctive $J=0$
polarization pattern. At the same time, the indication comes
invariably from low- and high-\pt\ data and is, therefore, more
``universal'' than the validity of the current factorized NLO
calculation, as established by the results of our high-\pt\ fits.
Furthermore, the $^1S_0^{[8]}$
polarization is zero at all perturbative orders and the
factorization prediction for the production via $^1S_0^{[8]}$ is,
obviously, unpolarized when resummed to all orders in any kind of
perturbative expansion, in agreement with data down to low \pt. This
leaves open the possibility that factorized calculations may
describe simultaneously high- and low-\pt\ data, if higher
perturbative orders improve the \pt\ description (specifically, by
reducing the steepness of the $^1S_0^{[8]}$ \pt\ distribution at low
\pt). Also in this case, polarization data show their power in
driving us towards encouraging indications on the reliability of
the NRQCD factorization framework.

Finally, we have also extrapolated the fitted cross sections and 
polarizations to very high \pt, providing predictions for future LHC 
measurements; the $^3S_1^{[8]}$ term seems to become more 
important, at least for the \upsThreeS, increasing the fraction of
transversely polarized mesons.

\bigskip

We congratulate our colleagues from the LHC collaborations for their 
high-quality quarkonium production measurements, 
without which this study could not have been made.
We acknowledge very interesting discussions with Geoff Bodwin and
Sergey Baranov. We are indebted to Mathias Butensch\"on and Bernd Kniehl,
who kindly gave us their NLO calculations of the short distance coefficients.
The work of P.F.\ is supported by FCT, Portugal, 
through the grant SFRH/BPD/42343/2007, while 
the work of V.K.\ is supported by FWF, Austria,
through the grant P24167-N16.


\begin{thebibliography}{00}
\parskip=5pt \parsep=0.pt \itemsep=0.pt

\bibitem{bib:Baier}
R.\ Baier and R.\ R\"uckl, 
Z. Phys. \textbf{C19} (1983) 251.\\
E.W.N. Glover, A.D. Martin, W.J. Stirling, 
Z. Phys. \textbf{C38} (1988) 473;
Erratum-ibid. \textbf{C49} (1991) 526.

\bibitem{bib:E789} 
M.H.\ Schub \emph{et al.} (E789 Coll.),
Phys. Rev. \textbf{D52} (1995) 1307; 
Erratum-ibid. \textbf{D53} (1996) 570.\\
P.L. McGaughey, 
Nucl. Phys. \textbf{A610} (1996) 394c.

\bibitem{bib:CDFpsisRunI} 
A. Sansoni \emph{et al.} (CDF Coll.), 
Nucl. Phys. \textbf{A610} (1996) 373c.\\
F.\ Abe \emph{et al.} (CDF Coll.), 
Phys. Rev. Lett. \textbf{79} (1997) 572.

\bibitem{bib:NRQCD} 
G.T.\ Bodwin, E.\ Braaten, G.P.\ Lepage, 
Phys. Rev. \textbf{D51} (1995) 1125; 
Erratum-ibid. \textbf{D55} (1997) 5853.

\bibitem{bib:QWGreportII} 
N.\ Brambilla  \emph{et al.} (QWG Coll.), 
Eur. Phys. J. \textbf{C71} (2011) 1534. 

\bibitem{bib:EPJC69} 
P.\ Faccioli, C.\ Louren\c{c}o, J.\ Seixas and H.K.\ W{\"o}hri,
Eur. Phys. J. \textbf{C69} (2010) 657.

\bibitem{bib:CDFpol1} 
T. Affolder \emph{et al.} (CDF Coll.),
Phys. Rev. Lett. \textbf{85} (2000) 2886.

\bibitem{bib:CDFpol2} 
A. Abulencia \emph{et al.} (CDF Coll.),
Phys. Rev. Lett. \textbf{99} (2007) 132001.

\bibitem{bib:CDFUpsPolRunI}
D.\ Acosta  \emph{et al.} (CDF Coll.),
Phys. Rev. Lett. \textbf{88} (2002) 161802.

\bibitem{bib:D0UpsPolRunII} 
V.M.\ Abazov \emph{et al.} (D0 Coll.),
Phys. Rev. Lett. \textbf{101} (2008) 182004.

\bibitem{bib:CDFUpsPolRunII} 
T.\ Aaltonen \emph{et al.} (CDF Coll.),
Phys. Rev. Lett. \textbf{108} (2012) 151802.

\bibitem{bib:CMSpsiPol} 
S.\ Chatrchyan \emph{et al.} (CMS Coll.),
Phys. Lett. \textbf{B727} (2013) 381.

\bibitem{bib:CMSUpsPol}
S.\ Chatrchyan \emph{et al.} (CMS Coll.),
Phys. Rev. Lett. \textbf{110} (2013) 081802.

\bibitem{bib:LHCbpsiPol}
R.\ Aaij \emph{et al.} (LHCb Coll.),
Eur. Phys. J. \textbf{C73} (2013) 2631.

\bibitem{bib:FT2C} 
P. Faccioli, C. Louren\c{c}o, J. Seixas and H.K. W\"ohri, 
Phys. Rev. Lett. \textbf{102} (2009) 151802.

\bibitem{bib:LTGen} 
P. Faccioli, C. Louren\c{c}o and J. Seixas,
Phys. Rev. Lett. \textbf{105} (2010) 061601.

\bibitem{bib:ImprovedQQbarPol} 
P. Faccioli, C. Louren\c{c}o and J. Seixas, 
Phys. Rev. \textbf{D81} (2010) 111502(R).

\bibitem{bib:BKmodel}
M. Butensch\"on, B.A. Kniehl,
Phys. Rev. Lett. \textbf{108} (2012) 172002
and private communication.

\bibitem{bib:CDFjpsi2005} 
D.\ Acosta \emph{et al.} (CDF Coll.),
Phys. Rev. \textbf{D71} (2005) 032001.

\bibitem{bib:BKNPB}
M. Butensch\"on, B.A. Kniehl,
Nucl. Phys. \textbf{B} (Proc. Suppl.) \textbf{222--224} (2012) 151.

\bibitem{bib:BKMPLA}
M. Butensch\"on, B.A. Kniehl,
Mod. Phys. Lett. A \textbf{28} (2013) 1350027.

\bibitem{bib:HERAb} 
I. Abt \emph{et al.} (HERA-B Coll.),
Eur. Phys. J. \textbf{C60} (2009) 525.

\bibitem{bib:ATLASjpsi2011} 
G.\ Aad \emph{et al.} (ATLAS Coll.),
Nucl. Phys. \textbf{B850} (2011) 387.

\bibitem{bib:CMSjpsi2012} 
S.\ Chatrchyan \emph{et al.} (CMS Coll.),
JHEP \textbf{02} (2012) 011.

\bibitem{bib:ATLASYnS2013}
G.\ Aad \emph{et al.} (ATLAS Coll.),
Phys. Rev. \textbf{D87} (2013) 052004.

\bibitem{bib:CMSYnS2013} 
CMS Coll.,
CMS-PAS-BPH-12-006.

\bibitem{bib:ATLASchi2013} 
ATLAS Coll.,
ATLAS-CONF-2013-095.

\bibitem{bib:ALICEpol} 
B.\ Abelev \emph{et al.} (ALICE Coll.), 
Phys. Rev. Lett. \textbf{108} (2012) 082001.

\bibitem{bib:LHCbpsip2012} 
R.\ Aaij \emph{et al.} (LHCb Coll.),
Eur. Phys. J. \textbf{C72} (2012) 2100. 

\bibitem{bib:LHCbYnS2012}
R.\ Aaij \emph{et al.} (CMS Coll.),
Eur. Phys. J. \textbf{C72} (2012) 2025.

\bibitem{bib:CSpartialNNLO}
J.P.\ Lansberg, 
Phys. Lett. \textbf{B679} (2009) 340.

\bibitem{bib:Chaofit} 
K.-T. Chao, Y.-Q. Ma, H.-S. Shao, K. Wang and Y.-J. Zhang, 
Phys. Rev. Lett. \textbf{108} (2012) 242004.

\bibitem{bib:Gaofit} 
B. Gong, L.-P. Wan, J.-X. Wang and H.-F. Zhang, 
Phys. Rev. Lett. \textbf{110} (2013) 042002.

\bibitem{bib:Schuler}
G.A.\ Schuler,
Int. J. Mod. Phys. \textbf{A12} (1997) 3951.

\bibitem{bib:Fleming} 
S.\ Fleming, I.Z.\ Rothstein and A.K.\ Leibovich,
Phys. Rev. \textbf{D64} (2001) 036002.

\bibitem{bib:chiPol} 
P.\ Faccioli, C.\ Louren\c{c}o, J.\ Seixas and H.K.\ W{\"o}hri,
Phys. Rev. \textbf{D83} (2011) 096001.

\end{thebibliography}
\end{document}